\documentclass{article}

\usepackage[nonatbib]{neurips_2024}
\usepackage{natbib}

\usepackage{neurips_2024}

\usepackage{bbding}
\usepackage{graphicx}
\usepackage{subcaption}
\usepackage{enumitem}
\usepackage{soul}
\usepackage{mathtools}
\usepackage{algorithm}
\usepackage{algorithmic}
\usepackage{xcolor}
\usepackage{colortbl}

\usepackage[utf8]{inputenc} % allow utf-8 input
\usepackage[T1]{fontenc}    % use 8-bit T1 fonts
\usepackage{hyperref}       % hyperlinks
\usepackage{url}            % simple URL typesetting
\usepackage{booktabs}       % professional-quality tables
\usepackage{amsfonts}       % blackboard math symbols
\usepackage{nicefrac}       % compact symbols for 1/2, etc.
\usepackage{microtype}      % microtypography
\usepackage{xcolor}         % colors

\title{Hybrid Quantum Downsampling Networks}

\author{%
 Yifeng Peng \\
  Stevens Institute of Technology 
  \And
  Xinyi Li \\
  Duke University 
  \And
 Zhiding Liang\\
Rensselaer Polytechnic Institute
  \AND
  Ying Wang\\
Stevens Institute of Technology 
}

\begin{document}

\maketitle

\begin{abstract}
 Classical max pooling plays a crucial role in reducing data dimensionality among various well-known deep learning models, yet it often leads to the loss of vital information. We proposed a novel hybrid quantum downsampling module (HQD), which is a noise-resilient algorithm. By integrating a substantial number of quantum bits (qubits), our approach ensures the key characteristics of the original image are maximally preserved within the local receptive field. Moreover, HQD provides unique advantages in the context of the noisy intermediate-scale quantum (NISQ) era. We introduce a unique quantum variational circuit in our design, utilizing rotating gates including $RX$, $RY$, $RZ$ gates, and the controlled-NOT (CNOT) gate to explore nonlinear characteristics. The results indicate that the network architectures incorporating the HQD module significantly outperform the classical structures with max pooling in CIFAR-10 and CIFAR-100 datasets. The accuracy of all tested models improved by an average of approximately $3\%$, with a maximum fluctuation of only $0.4\%$ under various quantum noise conditions.
\end{abstract}

\begin{figure}[ht]

\begin{center}
\centering
\centerline{\includegraphics[width=0.9\textwidth]{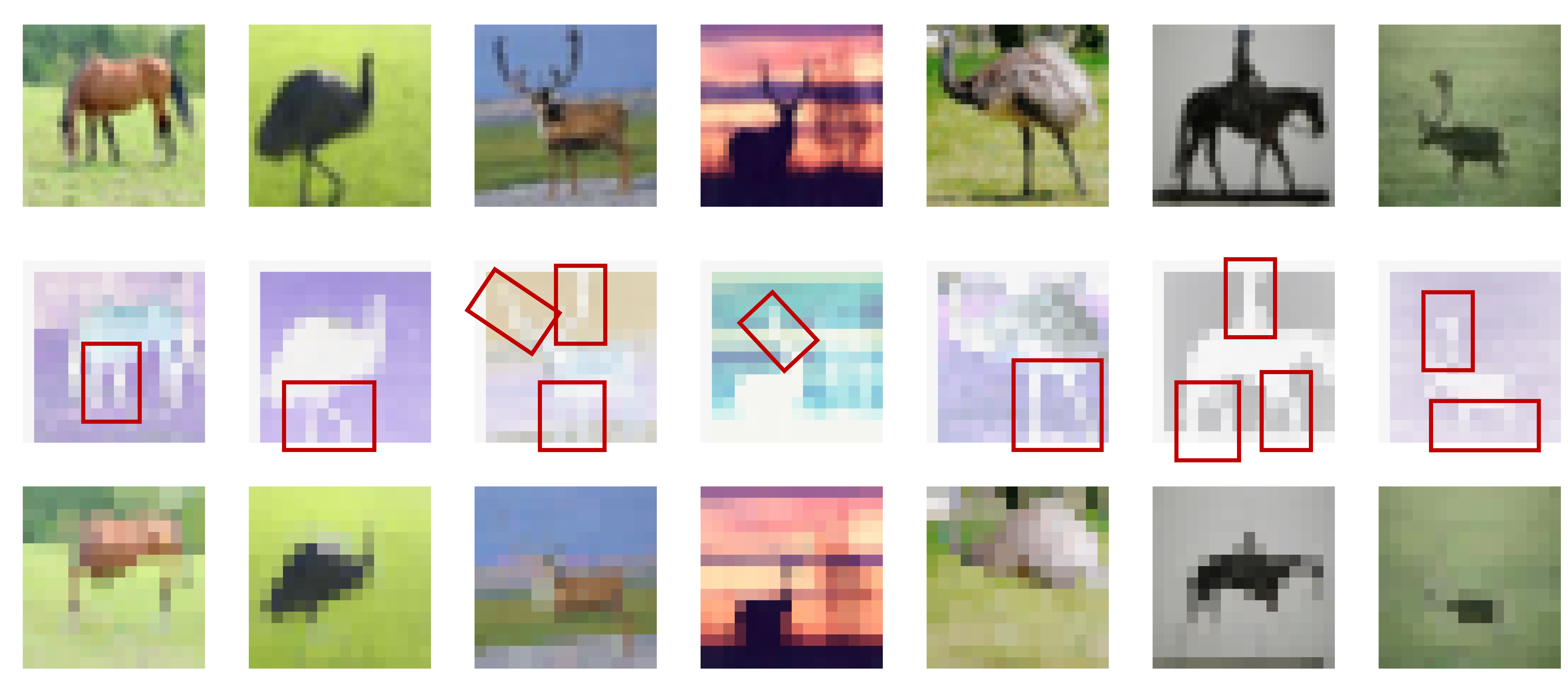}}
\caption{Visualization of images after different downsampling methods on \textbf{CIFAR-10}. Top: \textbf{Original images}. Middle: \textbf{Quantum Downsampling} (Ours). Bottom: \textbf{Max pooling}. Both are with kernel size to be $3 \times 3$, stride to be $2$, and padding to be $1$. More details are provided in the supplementary material as the space limitation.}
\label{Cifarviso}
\end{center}
\end{figure}

\section{Introduction}
\label{sec:intro}

%\textbf{Pooling $->$  Quantum Computing $->$ Quantum Computer Vision $->$ Hybrid Quantum Structure  $->$ Hybrid Pooling Strategy}

Max Pooling is a subsampling technique commonly used in deep learning, especially in convolutional neural networks (CNNs) \cite{NIPS1989_53c3bce6, CNN}. It is often used to reduce the spatial size of data and computational complexity and increase the receptive field of neurons in subsequent layers. Although the maximum value in max pooling can carry salient features, the features that are not selected during the pooling process are discarded, resulting in a loss of information \cite{maxpoolingavepooing}. 

Recently, there has been a growing interest in applying quantum computing in computer vision \cite{Birdal_2021_CVPR, Golyanik_2020_CVPR,melicvpr2022, Cong2019QuantumCNN, Zhang_2023_CVPR, Farina_2023_CVPR, Silver_2023_ICCV}. These quantum algorithms not only hold the promise of tackling intractable problems more effectively but also unveil new patterns and insights within large datasets that are currently beyond the reach of classical computational methods. The quantum-inspired and quantum-enabled algorithms have been adopted in various tasks, including machine learning (ML) \cite{Cerezo2022ChallengesAO, caro2022generalization}, natural language processing (NLP) \cite{li-etal-2019-cnm}, computer vision (CV) \cite{Tang_2022_CVPR, Golyanik_2020_CVPR,melicvpr2022, Benkner2021QMatchIS}, and multimodal analysis \cite{Gkoumas2021QuantumCM, Li2021QuantuminspiredNN}.

%The training method for the QNNs was proposed in \cite{NIPS2003_50525975}. 

 %Impressive advances both in quantum computing hardware and algorithms have been demonstrated over the last thirty years \cite{Patrickquantumbigdata, Lloyd2014QuantumPCA, PhysRevA.94.032329, Li2018QuantumAnnealing}

%Many quantum methods have been proposed for image processing \cite{ Dendukuri2018ImagePI, Yan2015ASO, Silver_2023_ICCV, Dendukuri2019DefiningQN, quantumimageprocessingcaraiman}.

%Quantum machine learning (QML) attempts to utilize this power of quantum computers to achieve computational speedups or better performance for machine learning tasks \cite{Biamonte2017QuantumML, Lloyd2016QuantumAlgorithmsData, PhysRevLett.113.130503, Huang:18}, and parameterized quantum circuits (PQCs) offer a promising path for quantum machine learning in the NISQ era.

However, in the noisy intermediate scale quantum (NISQ) era \cite{McClean_2016, Benedetti_2019, NISQ}, quantum computers cannot perform complex quantum algorithms for practical applications. Qubits and their operations are prone to errors due to quantum decoherence and other quantum noise. High error rates in qubit operations and quantum gate functions pose a major challenge to the accuracy of quantum algorithms. Quantum noise can lead to errors and information loss, which is particularly detrimental for quantum neural networks that require high precision.
%Every quantum gate operation has the potential to introduce errors because real-world quantum hardware is not perfect. Reducing the number of quantum gates reduces the overall error probability introduced by the operation, thereby increasing the overall accuracy of the quantum algorithm.

%In the realm of quantum neural networks (QNNs), the requisite one-to-one correspondence between qubits and input variables imposes a substantial constraint, given the current stage of quantum computing technology, which limits the applicability of QNNs for solving intricate real-world problems. This has steered the scientific inquiry towards the domain of hybrid quantum-classical neural network architectures, which are postulated as a more pragmatic approach in contrast to the purely quantum computational frameworks. Hybrid quantum neural networks also provide a platform for the transition between quantum and classical computing.

The current limitations of quantum computing technology have shifted scientific focus towards hybrid quantum-classical neural network architectures \cite{LaszloGyongyosiandSandorImreAsurveyonquantumcomputingtechnology, PerdomoOrtiz2017OpportunitiesAC, Yang_2022_CVPR, Doan2022AHQ, Bravyi2022hybridquantum, Domingo2023, Hybrid_HT_CNN, Bhatia_2023_CVPR}. These are viewed as a more pragmatic approach compared to the purely quantum computational frameworks due to the relatively higher noise immunities and the implementability of hybrid quantum neural networks.

Inspired by hybrid quantum architecture, we propose a hybrid quantum downsampling module (HQD), which overcomes the limitations of the classical max pooling method in existing classical deep learning models. HQD module can be applied to any state-of-the-art (SOTA) deep learning model as an alternative to max pooling to explore deeper nonlinear relationships between features, as shown in Fig. \ref{Cifarviso}. We use red circles to highlight the features that are easily overlooked in max pooling. Quantum computing brings us a wonderful new perspective on the world.

Within the existing framework of classical computing, no known algorithm is capable of simulating the behavior of a quantum computer \cite{preskill2018quantum}. This study unveils the distinctive features observable in the quantum realm by employing quantum variational circuits. Different from the feature maps seen by classical max pooling, the quantum downsampling we propose shows us a completely different computer vision in the quantum field in Fig. \ref{Cifarviso}. This process is fundamentally a downsampling procedure, markedly divergent from that of classical computers, capturing a richer array of details. We believe this will play a milestone role in the development of quantum computer vision in the future.

\section{Related Work}
\label{relatedwork}

In this section, we mainly focus on prior work related to quantum computing, quantum computer vision, and hybrid quantum-classical architecture.

\textbf{Quantum Computing.} Quantum computing has experienced a meteoric rise as a focal point of interdisciplinary research, with significant strides being made in the development of quantum hardware \cite{de2021materials, Kandala2017,wangisscc2023}. Parallel to these hardware advancements, the field of quantum algorithms\cite{Zhang_2023_CVPR, Huang_2023, Domingo2023} has seen a surge in activity.

\textbf{Quantum Computer Vision.} Quantum computer vision (QCV) is an exciting and evolving field with the potential to revolutionize how machines interpret visual data. Quantum approaches have been identified as catalysts for accelerating data processing capabilities and enhancing the performance of analytical models. Over the past few years, researchers have proposed several quantum-based algorithms aimed at addressing various computer vision tasks such as shape matching \cite{Golyanik_2020_CVPR,melicvpr2022, Benkner2021QMatchIS}, object tracking \cite{quboli2020, Zaech2022AdiabaticQC}, point triangulation \cite{Doan2022AHQ}, motion segmentation \cite{eccv2022quantummotion} and image classification \cite{Zhang_2023_CVPR}, multi-model fitting \cite{Farina_2023_CVPR}, image generation \cite{Silver_2023_ICCV}, point set registration \cite{melicvpr2022}, permutation synchronization \cite{Birdal_2021_CVPR}, and shape-matching \cite{Benkner2021QMatchIS}.

\textbf{Hybrid Quantum-Classical Architecture.} Based on the advantages of hybrid quantum-classical architecture \cite{LaszloGyongyosiandSandorImreAsurveyonquantumcomputingtechnology, PerdomoOrtiz2017OpportunitiesAC, Doan2022AHQ}, related work has gradually begun to appear in the field of computer vision in recent years \cite{Yang_2022_CVPR}.
%(CVPR 2022) 
%Doan  \cite{Doan2022AHQ} proposed a hybrid quantum-classical algorithm, which solves a series of integer programming problems and ultimately obtains a global solution or error bound.
Bravyi  \cite{Bravyi2022hybridquantum} formulated a variation of the Quantum Approximate Optimization Algorithm (QAOA) using qudits which is applicable to non-binary combinatorial optimization. Domingo  \cite{Domingo2023} proposed a hybrid quantum-classical three-dimensional convolutional neural network where one or more convolutional layers are replaced by quantum convolutional layers. Pan  \cite{Hybrid_HT_CNN} introduced an innovative approach to hybrid quantum-classical computing by integrating a novel Hadamard Transform (HT)-based layer into neural networks. Bhatia  \cite{Bhatia_2023_CVPR} proposed a new quantum-hybrid method for solving the problem of multiple matching of non-rigidly deformed 3D shapes. 

\section{Methodology}

% \subsection{Max Pooling}
% In max pooling, the size of the input image $I_M [\cdot]$ is $H \times W$ and the dimension of input image after max pooling $\tilde{I_M}[\cdot]$ is shown as below: 

% \begin{align}
% H_{\text{out}} &= \left\lfloor \frac{H_{\text{in}} - K + 2 \times p}{s} + 1 \right\rfloor \notag \\
% W_{\text{out}} &= \left\lfloor \frac{W_{\text{in}} - K + 2 \times p}{s} + 1 \right\rfloor ,
% \label{EQ-1111}
% \end{align}
% where $H_{in}$ and $W_{in}$ are the dimensions of the input image. And $\left\lfloor \cdot \right\rfloor$ represents the flooring operation. The max pooling can be expressed as below:
    
% \begin{equation}
%    \tilde{I_M}(i,j) = \max_{a=0}^{K-1} \max_{b=0}^{K-1} I_M(s \cdot i + a - p, s \cdot j + b - p),
% \end{equation}
% where $\tilde{I_M}(i,j)$ is the pixel value in the position of $(i,j)$ in the output feature image. $K$ is the kernel size, $s$ is the stride, and $p$ is the padding. $a$ is the index of the row in the pooling window, and $b$ is the index of the column in the pooling window.

\subsection{Quantum Downsampling}

\begin{figure}[h]

\begin{center}
\centerline{\includegraphics[width=0.9\textwidth]{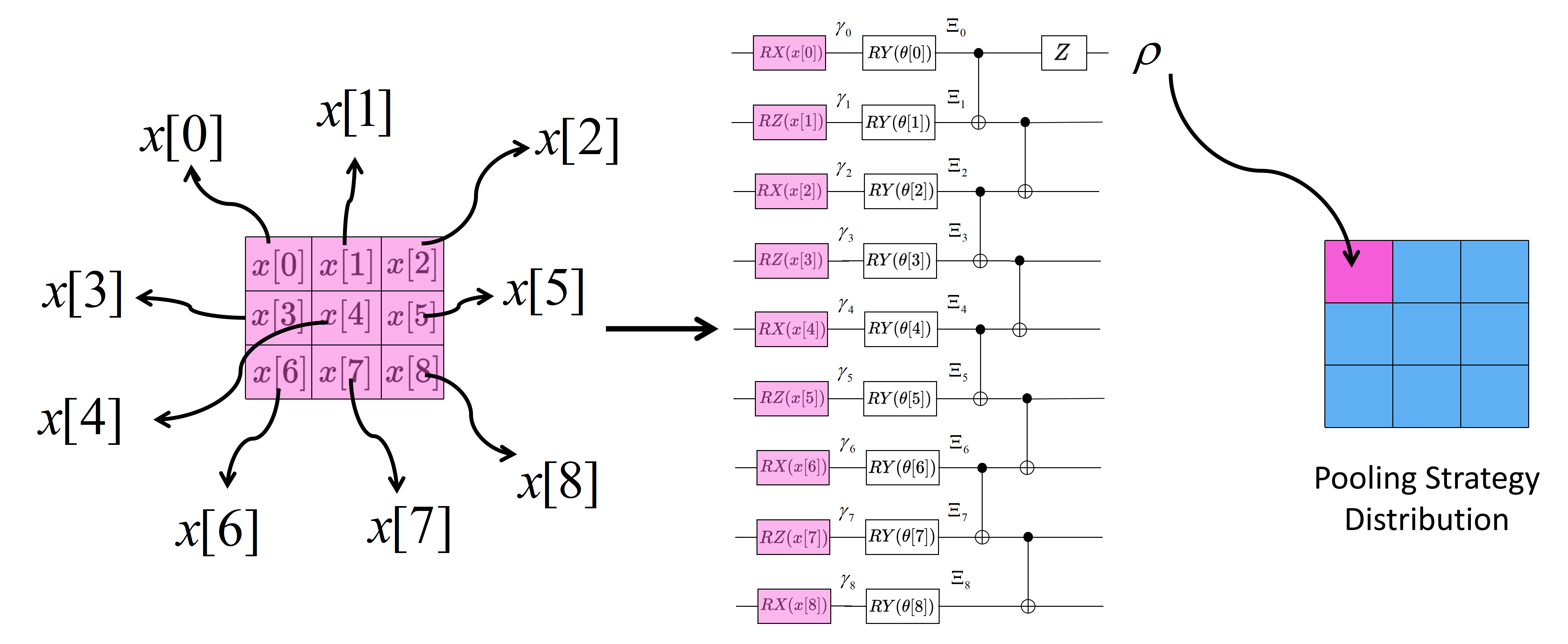}}
\caption{Structure of $3 \times 3$ quantum downsampling: it shows the $3 \times 3$ pooling window, extracts $9$ eigenvalues, and inputs them into the quantum variational circuit to obtain an output. \textbf{Pink} color indicates the use of \textbf{quantum downsampling}. \textbf{Pink} color in the quantum variational circuit $x[0]-x[8]$ is the value from the input image, and $\theta$ is the parameters to be optimized and $\gamma$, $\Xi$ are values in quantum variational circuit. \textbf{Blue} color indicates the use of \textbf{classical max pooling}. The pooling strategy distribution is determined by $\ell$. The $2 \times 2$ quantum downsampling structure is placed in \textbf{Appendix} Fig. \ref{quantumpooling2X2}.}
\label{quantumpooling3X3}
\vspace{-20pt}
\end{center}

\end{figure}

This section presents the framework of the quantum downsampling method with kernel size to be $2 \times 2 $ and $ 3 \times 3$ with $4$ qubits and $9$ qubits, respectively. Since the internal max-pooling of most SOTA models uses a $3 \times 3$ kernel size, only the $3 \times 3$ HQD module is discussed in this section, and the relevant parts of $2 \times 2$ are placed in supplementary materials.
We proposed the forward propagation process of quantum variational circuits throughout the network as well as the gradient descent process.

In Fig. \ref{quantumpooling3X3}, the design of the HQD quantum variational circuit consists of $RX$, $RY$, $RZ$ gates, and CNOT gates, among which Pauli-Z is used to measure the final output. For $RX$ and $RZ$, the rotation amplitude is determined by the value of the input feature map $x$, but for RY, the rotation amplitude is determined by an optimizable parameter $\theta$.

\begin{figure}[h]
\begin{center}
\centerline{\includegraphics[width=1\textwidth]{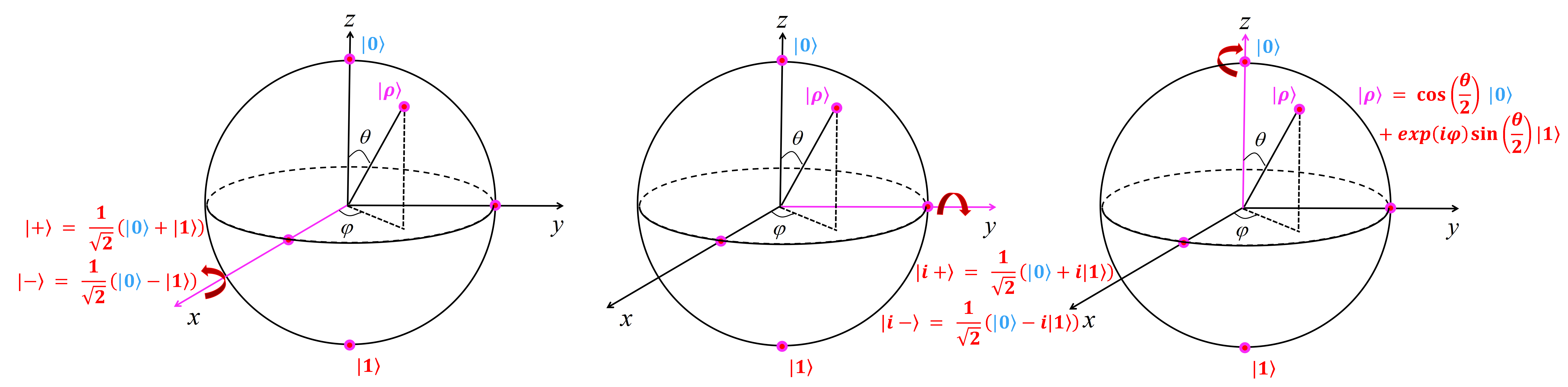}}
\caption{Bloch sphere representation of the quantum state $\left | \rho  \right \rangle $ during the quantum downsampling. Rotation logic gates $RX$ gate (Left), $RY$ gate (Middle), and $RZ$ gate (Right) for single qubit.}
\label{FIG-AA}
\vspace{-20pt}
\end{center}
\end{figure}

As for the quantum variational circuit of HQD, part of it is an alternating design of $RX$ and $RZ$. The $RX$ and $RZ$ gates are rotation operations of the qubit state on the Bloch ball, rotating around the x-axis and z-axis shown in Fig. \ref{FIG-AA}, respectively. By using these two rotations alternately, quantum states can be explored in different dimensions, increasing the ability of quantum circuits to cover Bloch spheres. The $RY$ gate provides rotation around the y-axis shown in Fig. \ref{FIG-AA}, further increasing the ability to explore the third dimension. This comprehensive rotation strategy helps generate richer and more diverse quantum states. Rotations in different axes have different sensitivities to different system errors, and using them alternately can help spread the effects of these errors.

By inserting a CNOT gate (used to generate entanglement between two qubits) between quantum gate operations (such as $RX$ and $RZ$ rotation), the entanglement between qubits can be effectively generated and controlled. The design strategy of alternately using $RX$ and $RZ$ structures with $RY$ gates increases the exploration capabilities, expression capabilities, and entanglement generation capabilities of quantum circuits, as well as providing error mitigation and flexibility.

The introduction of optimizable parameters $\theta$ into quantum circuits, especially the use of these parameters in $RY$ gates, makes the circuit parametric. This means that the output of the circuit can be optimized by adjusting these parameters to minimize a certain loss function or achieve a certain computational goal. The $RY$ gate directly adjusts the probability amplitude of the qubit, affecting its probability of being in the $\left |0 \right \rangle $and $\left | 1 \right \rangle $ states. It is very suitable for controlling the degree of superposition of quantum states and realizing conversion between quantum states.

% $RY$ gates (rotation about the y-axis) provide a more direct way to control the superposition and probability amplitude of quantum states in some cases. Because in the Bloch ball representation, rotation around the y-axis directly affects the superposition of the $\left |0 \right \rangle $ and $\left | 1 \right \rangle $ states of the qubit, this may be more efficient or intuitive for some algorithms. In contrast, the rotation of the $RX$ and $RZ$ gates affects the real and imaginary parts of the quantum state, respectively, which is not the most direct adjustment method when dealing with the problem.

A unitary operator $U(\cdot)$ for the quantum circuit is defined as

\begin{equation}
U = \exp \left( {i\alpha } \right)\left[ {I\cos \frac{\theta }{2} - i\left( {\vec \sigma .\hat n} \right)\sin \frac{\theta }{2}} \right] \times \exp \left[ {-i\frac{\theta }{2}\left( {\vec \sigma .\hat n} \right)} \right] 
\label{EQ-1},
\end{equation}

where ${\vec \sigma .\hat n} = \sigma_xn_x + \sigma_yn_y + \sigma_zn_z$. The operator $U(\cdot)$ can readily be obtained in simplified form as illustrated in Eq. (\ref{EQ-44}). The variables $\varpi, \chi, \Lambda$, and $\nu$ are all real numbers.

\begin{equation}
U = \left[ \begin{array}{cc} e^{i\left( \varpi - \frac{\chi}{2} - \frac{\Lambda}{2} \right)\cos \frac{\nu}{2}} & -e^{i\left( \varpi - \frac{\chi}{2} + \frac{\Lambda}{2} \right)\sin \frac{\nu}{2}} \\ e^{i\left( \varpi + \frac{\chi}{2} - \frac{\Lambda}{2} \right)\sin \frac{\nu}{2}} & e^{i\left( \varpi + \frac{\chi}{2} + \frac{\Lambda}{2} \right)\cos \frac{\nu}{2}} \end{array} \right]. \label{EQ-44}
\end{equation}

Let $RX(\theta)$ be the rotation operator that rotates the qubit through angle $\theta {\rm(radians)}$ around the $X-$axis. It is defined as
\begin{equation}
RX\left( \theta  \right) = \left[ {\begin{array}{*{20}c}
   {\cos \frac{\theta }{2}} & { - i\sin \frac{\theta }{2}}  \\
   { - i\sin \frac{\theta }{2}} & {\cos \frac{\theta }{2}}  \\
\end{array}} \right] .
\label{EQ-2}
\end{equation}
Applied to any arbitrary pixel $x[k]$, the rotation operator $RX(x[k])$ yields an output as illustrated in Eq. (\ref{EQ-3}):

\begin{equation}
x[k] \xrightarrow[]{RX(x[k])} \gamma = \cos \frac{\theta}{2}\left( \left| 0 \right\rangle \langle \left. 0 \right| + \left| 1 \right\rangle \langle \left. 1 \right| \right)x[k] - i\sin \frac{\theta}{2}\left( \left| 1 \right\rangle \langle \left. 0 \right| + \left| 0 \right\rangle \langle \left. 1 \right| \right) x[k].
\label{EQ-3}
\end{equation}

The output of $RX(\theta)$ and $RZ(\theta)$ operations are then applied to the quantum gate $RY(\theta)$, defined as
\begin{equation}
RY\left( \theta  \right) = \left[ {\begin{array}{*{20}c}
   {\cos \frac{\theta }{2}} & { - \sin \frac{\theta }{2}}  \\
   {\sin \frac{\theta }{2}} & {\cos \frac{\theta }{2}}  \\
\end{array}} \right] .
\label{EQ-4}
\end{equation}
The output $\gamma$ is fed to the input of the gate $RY(\theta)$, which yields an output defined as depicted in Eq. (\ref{EQ-5}):

\begin{equation}
\begin{split}
    & \gamma_j \hspace{0.2cm} \xrightarrow[]{RY(\gamma_j)} \Xi_j =  \hspace{0.2cm}  \cos \frac{\theta }{2}\left( {\left| 0 \right.\rangle \langle \left. 0 \right| + \left| 1 \right.\rangle \langle \left. 1 \right|} \right)\gamma_j \\
    & + \sin \frac{\theta }{2}\left( {\left| 1 \right.\rangle \langle \left. 0 \right| - \left| 0 \right.\rangle \langle \left. 1 \right|} \right) \gamma_j; j \in \{0, 1, \cdot\cdot\cdot, J\} .
\end{split}
    \label{EQ-5}
\end{equation}

As shown in Fig. \ref{quantumpooling3X3}, the $RZ(\cdot)$ is defined as
\begin{equation}
    RZ(\theta) = e^{\frac{-i\theta Z}{2} } = \cos \left ( \frac{\theta}{2}  \right )I-i\sin \left ( \frac{\theta}{2}  \right )  .
\label{RZ}
\end{equation}

In Fig. \ref{quantumpooling3X3}, the index $j$ represents the index of the wire in the quantum circuit. $J = 3$ for $2 \times 2$ quantum downsampling kernel and $J = 8$ for $3 \times 3$ quantum downsampling kernel. The quantum gate $Z$, as illustrated in Fig. \ref{quantumpooling3X3} corresponds to the Pauli-Z gate and is represented by
\begin{equation}
Z\left( \Xi  \right) = \left[ {\begin{array}{*{20}c}
   1 & 0  \\
   0 & { - 1}  \\
\end{array}} \right]\Xi .
\label{EQ-6}
\end{equation}
The corresponding output $\rho$ can then be evaluated as
\begin{equation}
\rho  = \left( {\left| 0 \right.\rangle \langle \left. 0 \right| - \left| 1 \right.\rangle \langle \left. 1 \right|} \right) \otimes \Xi
\label{EQ-7},
\end{equation}
where $\otimes$ denotes the tensor product operation.
$\Xi$ propagates through the controlled NOT gate. The control not gate, $N_C(\cdot)$ is defined as
\begin{equation}
N_C \left( \theta  \right) = \exp \left( {i\frac{\pi }{4}\left( {I - Z} \right)\left( {I - X} \right)} \right)
\label{EQ-13}.
\end{equation}
It can be easily shown that
\begin{equation}
N_C \left( \theta  \right)N_C^{\dag} \left( \theta  \right) = I ,
\label{EQ-15}
\end{equation}
where $I$ represents the identity matrix.

\subsection{{Quantum Measurement of  $|\rho\rangle$ for Quantum Downsampling}}
Let $\mathcal{O}_l$ be the measurement operator acting on the quantum state $|\rho\rangle$. The index $l$ depicts the measurement outcomes that may occur in the process. For the given quantum state $|\rho\rangle$, the probability that result $l$ occurs after the quantum measurement is given by
\begin{equation}
    {\rm P}_{r} \left( l \right) = \langle \rho  {|\mathcal{O}_l^\dag  \mathcal{O}_l \left| {\rho } \right.} \rangle.
    \label{EQ-9}
\end{equation}
The quantum state after the measurement can then readily be estimated as
\begin{equation}
    \left| {\rho^l \rangle } \right. = \frac{{\mathcal{O}_l \left| {\rho } \right.\rangle }}{{\sqrt {\langle \rho  {| \mathcal{O}_l^\dag  \mathcal{O}_l \left| {\rho } \right.} \rangle } }}
    \label{EQ-17}.
\end{equation}
Since, by the law of total probability, $\sum\limits_l {{\rm P}_{r} \left( l \right)}  = \sum\limits_l {\langle \rho | {\mathcal{O}_l^\dag  \mathcal{O}_l \left| {\rho } \right.} \rangle  = 1}$, it can be readily shown that the measurement operators $\{\mathcal{O}_l\}; \forall l$ satisfy the completeness equation as
\begin{equation}
    \sum\limits_l {\mathcal{O}_l^\dag  \mathcal{O}_l  = I}.
    \label{EQ-18}
\end{equation}
Moreover, for any quantum state $\left| \rho\rangle \right.$, the probability of obtaining result $l$ can then be written as
\begin{equation}
    \begin{split}
& {\rm P}_{r} \left( l \right) = \sum\limits_{j = 1}^{J } {{\rm tr}\left( {\mathcal{O}_l^\dag  \mathcal{O}_l \sum\limits_{j = 1}^{J } {\left. {{\rm P}_{r} } \right|\rho \rangle \langle \left. {\rho } \right|} } \right)},  \\
    \end{split}
\end{equation}
where ${\rm tr(\cdot)}$ is the trace operator.
The quantum state representation during the quantum downsampling process is illustrated in Fig. \ref{FIG-AA}.

$RX$, $RY$ can bring changes in probability amplitude in Eq. (\ref{EQ-2}) and Eq. (\ref{EQ-4}), while $RZ$ only changes in phase in Eq. (\ref{RZ}). Using these three operations together allows quantum states to move freely throughout the Bloch sphere or in the Hilbert space. By mixing three different quantum gate operations, more in-depth image nonlinear relationships can be explored, and therefore, the hybrid quantum learning model can perform the classification operation with improved performance.

\subsection{Quantum Downsampling Gradient Descent}
The learning of the proposed HQD is represented by the pooling gradient, as shown in Eq. (\ref{EQ-11111})
\begin{equation}
\frac{\mathrm{d} \left \langle \mathcal{O}_l \right \rangle }{\mathrm{d} {\theta_{i} } } \approx \frac{1}{2} \left [ \left \langle \mathcal{O}_l \right   \rangle \left ( \theta_i + \pi /2 \right ) - \left \langle \mathcal{O}_l \right   \rangle \left ( \theta_i - \pi /2 \right ) \right ] 
\label{EQ-11111}.
\end{equation}
During the training of the hybrid quantum model, the updated rule of the quantum downsampling gradient descent can be expressed as
\begin{equation}
\Theta _{t + 1}  \leftarrow \Theta _t  - \eta \frac{{d\langle \mathcal{O}_l \rangle }}{{d\theta _i }}
\label{EQ-BBB},
\end{equation}
where $\eta$ represents the learning rate.

To quantify the performance of the hybrid quantum-classical approach, we define the loss function as illustrated in Eq. (\ref{EQ-ZYX}) $\mathcal{L}$:
\begin{equation}
    \mathcal{L} =-\frac{1}{N}  {\textstyle \sum_{i= 1}^{N}}  {\textstyle \sum_{c=1}^{C}}y_{ic}\log_{}{\hat{y} _{ic}}, 
    \label{EQ-ZYX}
\end{equation}
where $N$ is the number of samples, $C$ is the number of classes, $y_{ic}$ is $1$ if the $i_{th}$ sample belongs to class $c$ and $0$ otherwise (one-hot encoding of the labels), and ${\hat{y} _{ic}}$ is the predicted probability (output of softmax in the last layer of the neural network) of the $i_{th}$ sample belonging to class $c$. It is to be noted that $\mathcal{L}$ reflects the sum of products of the expected value and log of the predicted value of all the pixels in the hybrid quantum-classical learning model.

\subsection{Whole Process of the HQD Module}
$\ell$ determines when to perform classical max pooling and when to perform the quantum downsampling process, as shown in the Algorithm. \ref{ALGO-1}. When it comes to quantum downsampling, the whole process is as follows:

\textbf{Extraction of Values.}
The value of each of the nine pixels within the $3 \times 3$ pooling window is extracted. The initial classical pixels $x[\cdot]$ are first processed through the quantum unitary operator $U(\cdot)$, and subsequently through the quantum gates $RX$, $RZ$ and $RY(\cdot)$ and Pauli-Z, respectively. 

\textbf{Application to the Quantum Variational Circuit.}
The extracted values are then applied to a quantum variational circuit. These circuits are characterized by a set of parameters that can be tuned or optimized during the process. The input from the pooling window (the extracted pixel values) is used to set or adjust these parameters.

\textbf{Obtaining Measurement Output.} After the quantum variational circuit processes the input data (the qubits corresponding to the $3 \times 3$ pooling window), a measurement operator $\mathcal{O}_l$ is operated on the quantum state $\rho$ of the system, which yields the output with probability ${\rm P}_r(l)$.
This measurement yields an output $\left| {\rho^l \rangle } \right.$, which is a processed value or set of values derived from the original $3 \times 3$ pixel data. In quantum computing, measurements typically collapse the superposed states of qubits into definite quantum states $|\rho\rangle$, which can then be interpreted or further processed.

\section{Quantum Noise}

To assess and enhance the resilience of quantum neural networks amidst disturbances, we introduced quantum errors in our experiments.

\textbf{Amplitude Damping.} The amplitude damping channel models the process where energy is lost to the surroundings, exemplified by a qubit transitioning from the excited state $\left | 1 \right \rangle$ to the ground state $\left | 0 \right \rangle$. This channel is depicted for individual qubits as
\begin{equation}
    \mathcal{C}\left ( \rho \right )  =  E_0 \rho E_0^{\dagger} + E_1 \rho E_1^{\dagger},
\end{equation}
where $E_0=\begin{pmatrix}
  1& 0\\
  0&\sqrt{1-\psi } 
\end{pmatrix}$ and $E_1=\begin{pmatrix}
  0& \sqrt{\psi}\\
  0&0 
\end{pmatrix}$. In addition, $\psi$ is the amplitude damping probability.

% In the HQD Network, each qubit is described by a quantum density matrix $\rho$, given by $\rho = \begin{pmatrix}
%  \rho_{00} & \rho_{01}\\
%  \rho_{10} & \rho_{11}
% \end{pmatrix}$
% , the effect of the amplitude damping channel on this density matrix $\rho$ is characterized as follows:
% \begin{equation}
%     \mathcal{C}_\psi   = \left [ \begin{pmatrix}
%  \rho_{00} & \rho_{01}\\
%  \rho_{10} & \rho_{11}
% \end{pmatrix} \right ] =\begin{pmatrix}
%  \rho_{00}+ \psi \rho_{11} & \sqrt{1-\psi } \rho_{01}\\ 
%  \sqrt{1-\psi } \rho_{10} &  (1-\psi \rho_{11})
% \end{pmatrix}.
% \end{equation}

 \textbf{Phase Flip Noise.} Given a phase flipping probability $\psi$, this type of noise operates by applying a Pauli $Z$ gate to the quantum state. This action flips the phase of the state $\left | 1 \right \rangle$ by introducing a negative sign, while the state $\left | 0 \right \rangle$ remains unaffected. It can be represented as
\begin{equation}
    \mathcal{N}(\rho) = (1-\psi) \rho + \psi Z\rho Z.
\end{equation}
% And in \cite{Nielsen_Chuang_2010}, the operation elements can be represented as:
% \begin{equation}
%     E_0 = \sqrt{\psi} I=\sqrt{\psi}\begin{bmatrix}
%   1&0 \\
%   0&1
% \end{bmatrix},
% \end{equation}
% and 
% \begin{equation}
%     E_1 = \sqrt{1-\psi} Z=\sqrt{1-\psi}\begin{bmatrix}
%   1&0 \\
%   0&-1
% \end{bmatrix},
% \end{equation}
% where $E_0$ represents the operation applied to the state $\left | 0 \right \rangle$, and $E_1$ denotes the operation targeting the state $\left | 1 \right \rangle$.

\textbf{Bit Flip Noise.} Bit flip noise transforms the state of a qubit from $\left | 0 \right \rangle$ to $\left | 1 \right \rangle$ or from $\left | 1 \right \rangle$ to $\left | 0 \right \rangle$ with a probability $\psi$, mirroring the effect of a classical bit flip error. Utilizing the Pauli-X operation, the updated state of $\rho$ is expressed as follows:
\begin{equation}
    \mathcal{N}(\rho) = (1-\psi) \rho +\psi X\rho X.
\end{equation}
% In \cite{Nielsen_Chuang_2010}, the elements of the operation are depicted as:
% \begin{equation}
%     E_0 = \sqrt{\psi} I=\sqrt{\psi}\begin{bmatrix}
%   1&0 \\
%   0&1
% \end{bmatrix}.
% \end{equation}
% and 
% \begin{equation}
%     E_1 = \sqrt{1-\psi} X=\sqrt{1-\psi}\begin{bmatrix}
%   0&1 \\
%   1&0
% \end{bmatrix}.
% \end{equation}

\textbf{Depolarizing Noise.} This noise models the information loss that occurs when a qubit interacts with its environment, potentially rendering the qubit's state entirely random with a given probability. Such interactions lead to a decline in the integrity of quantum information. The state of the qubit under depolarizing noise is shown in \cite{Nielsen_Chuang_2010} as
\begin{equation}
    \mathcal{N} \left ( \rho \right ) =\frac{\psi I}{2} +\left ( 1 - \psi\right ) \rho
    \label{depolar}.
\end{equation}
For the operator-sum representation, we can expand the mixed state $I/2$ as
\begin{equation}
    \frac{I}{2} =\frac{\rho+X\rho X+Y\rho Y+Z\rho Z}{4} ,
\end{equation}
and replace the $I/2$ in Eq. \ref{depolar} and it can be derived as
\begin{equation}
    \mathcal{N}\left ( \rho \right )   =  \left ( 1-\frac{3}{4} \psi \right )\rho + \frac{X\rho X+Y\rho Y+Z\rho Z}{4} \psi.
    \label{223}
\end{equation}
With the probability $\psi$, we can make Eq. \ref{223} as
\begin{equation}
    \mathcal{N}\left ( \rho \right )   =  \left ( 1- \psi \right )\rho + \frac{ \psi}{4} (X\rho X+Y\rho Y+Z\rho Z).
    \label{22}
\end{equation}

Depolarizing noise causes the state of a qubit to transition to a completely mixed state with probability $\psi$, while it remains unchanged with a of $1 - \psi$.

\section{Experimental Evaluation}
\label{experimentalevaluation}
\subsection{Hyperparameter Selection and Computing Resources}
The CIFAR-10 dataset is normalized with the mean (0.4914, 0.4822, 0.4465) and the standard deviation (0.247, 0.243, 0.261). We adopt the stochastic gradient descent (SGD) \cite{Robbins1951ASA} optimizer with momentum to be $0.9$ and weight decay to be $5e^{-4}$. The initial learning rate $\eta$ is set to $1e^{-3}$ and decreased by a factor of $10$ every $200$ epochs. The batch size is $128$, and the training epochs are $500$. We set up the simulation of a quantum variational circuit based on Pennylane\cite{Pennylane}. All the experiments are performed on an NVIDIA $3060$ GPU and $48$ GB RAM and an NVIDIA $4080$ GPU and $64$ GB RAM. The CIFAR-100 dataset is normalized with the mean (0.4914, 0.4822, 0.4465) and the standard deviation (0.247, 0.243, 0.261). We adopt the stochastic gradient descent (SGD) \cite{Robbins1951ASA} optimizer with momentum to be $0.9$ and weight decay to be $3e^{-4}$. The initial learning rate $\eta$ is set to $1e^{-2}$ and decreased by a factor of $5$ every $30$ epochs. The batch size is $128$, and the training epochs are $100$. For the experiment under quantum noise, including amplitude damping noise, phase flip noise, bit flip noise, and depolarizing noise, we ran the experiment $10$ times, and the noise factor $\psi$ is set to be $0.5$. More details of the structure of HQD models are in \textbf{Appendix} Section \ref{secvgg} to Section \ref{secxt}. In addition, the line plots and scatter plots of Table. \ref{CIFAR100-2}, Table. \ref{CIFAR100-3} and Table. \ref{CIFAR100} are provided in \textbf{Appendix} Fig.\ref{scatter15} and Fig.\ref{paramacc}.

\subsection{Experimental Results}

\subsubsection{CIFAR-10.}
\label{cifar100results}
From Table \ref{cifar10acc}, it is evident that HQD has notably enhanced the performance of classical models, albeit with varying degrees of improvement across different architectures. Notably, ResNet-18 benefits the most from the HQD module, witnessing an improvement rate of $3.8\%$. However, when $\ell=300$, the enhancement becomes negligible. On the other hand, the least improvement is observed in the Xception and Attention-56 models.
\begin{table*}[ht]
\centering
\caption{Experimental Results (Accuracy $(\%)$) on CIFAR-10 dataset.}
\resizebox{0.95\textwidth}{!}{
%\fontsize{6pt}{8pt}\selectfont
\setlength{\tabcolsep}{6pt} %
\begin{tabular}{ccccc}

\toprule
              Backbone     &  Original & HQD ($\ell = 100$) & HQD ($\ell = 300$) & Parameters (M) \\ 
                    \midrule
ResNet-18 \cite{Resnet} &     $75.50$   &     $ \textbf{78.36}\pm \textbf{0.32}  $     &      $75.81 \pm 0.29 $ & $11.22^\dagger $ \\ 
    Attention-56 \cite{Attention}  &   $71.86$  &     $72.54 \pm 0.19 $     &            $\textbf{72.74} \pm \textbf{0.24} $  &  $55.70^\dagger$\\  
Xception \cite{Chollet2016XceptionDL}   &  $79.29$  &     $79.77 \pm 0.08$      &      $\textbf{79.87} \pm \textbf{0.06}$  & $20.83^\dagger$\\ 
SE-ResNet-18 \cite{Hu2017SqueezeandExcitationN}  &   $76.36$ 
    &     $\textbf{78.19} \pm \textbf{0.15} $      &      $76.68 \pm 0.11$  & $11.20^\dagger$\\ 
SqueezeNet \cite{Iandola2016SqueezeNetAA}  &   $78.86$ 
    &     $\textbf{80.01} \pm \textbf{0.37}$     &      $79.01 \pm 0.28$  & $0.73^\dagger$\\ 
NASNet-A \cite{Zoph2017LearningTA} &   $81.15$ 
    &     $83.33 \pm 0.25$     &      $\textbf{83.41} \pm \textbf{0.29}$   & $5.13^\dagger$ \\ 
% ResNext-50 \cite{Xie2016AggregatedRT}  &   $65.47$ 
%    &      $65.82$  &     $41.30$  & \\
\bottomrule
\multicolumn{4}{l}{ $\dagger$ indicates $9$ more parameters in the HQD variational circuit.}\\
\multicolumn{4}{l}{\textbf{Bold} areas mean better performance under the same model between the two $\ell$s.}

\end{tabular}
}
\label{cifar10acc}
\end{table*}

\subsubsection{CIFAR-100.} This section delves deeper into our analysis by employing the more complex CIFAR-100 dataset. CIFAR-100 contains $60,000$ $32 \times 32$ pixel color images, CIFAR-100 contains $100 $ categories, and each category contains $600$ images. We tested four $\ell$ values $(50,100,500,1000)$ to explore the impact of different participation levels of the HQD module on the classic model. Based on the experimental results shown in Table. \ref{CIFAR100-2}, Table. \ref{CIFAR100-3} and Table. \ref{CIFAR100}, the best top-1 error and top-5 error of most models are concentrated in the HQD models with $\ell = 100$. It is observed that both extremes of $\ell$—too small, indicating minimal HQD participation in the pooling process, and too large, denoting extensive HQD involvement—are detrimental to HQD's effectiveness in enhancing the performance of classical models. Max pooling simplifies the feature map by extracting the maximum value within a local area, which is effective but may lead to information loss. HQD seeks to minimize information loss with intricate processing. However, overdependence on HQD can overly complicate the model's extracted features, impairing performance. Striking an optimal balance is crucial to ensure effective feature extraction and mitigate overfitting risks.
\begin{table*}[ht]
\centering
\caption{Experimental Results on CIFAR-100 for ResNet-50 and ResNeXt-50.}
\setlength{\tabcolsep}{4pt} 
%\fontsize{10pt}{10pt}\selectfont
\resizebox{0.9\textwidth}{!}{
\begin{tabular}{lccc}

\toprule
              Method     & Top-1 err. ($\%$) $(\downarrow)$ & Top-5 err. ($\%$)$(\downarrow)$ & Parameters (M) $(\downarrow)$\\ 
                    \midrule
ResNet-50 \cite{Resnet} &     $49.88$     &      $23.79$     &     $23.71 $\\ 
\midrule  
HQD ResNet-50 ($\ell=50$)  &     $ 50.28 \pm 0.21$     &      $24.53 \pm 0.18$  &     $23.71 ^\dagger $ \\  
HQD ResNet-50 ($\ell=100$)  &     $ \cellcolor{red!50} 47.89 \pm 0.24$     &      $\cellcolor{red!50} 22.37 \pm 0.22 $  &     $23.71 ^\dagger  $ \\  
HQD ResNet-50 ($\ell=500$)  &     $ 51.59 \pm 0.19$     &      $25.19 \pm 0.18 $  &     $23.71 ^\dagger $ \\  
HQD ResNet-50 ($\ell=1000$)  &     $ \cellcolor{yellow!40} 49.55 \pm 0.17 $     &      $\cellcolor{yellow!40}23.26 \pm 0.21$  &     $23.71 ^\dagger $ \\  
 \midrule
ResNeXt-50 \cite{Xie2016AggregatedRT}  &   $53.18$ 
    &      $26.37$  &     $14.79$ \\
\midrule  
HQD ResNeXt-50 ($\ell=50$)  &     $\cellcolor{yellow!40}52.93 \pm 0.43$     &      $26.91 \pm 0.29 $  &     $14.79 ^\dagger $ \\ 
HQD ResNeXt-50 ($\ell=100$)  &     $\cellcolor{red!50} 52.81 \pm 0.27  $     &      $ \cellcolor{red!50}26.00 \pm0.40 $  &     $14.79 ^\dagger $ \\ 
HQD ResNeXt-50  ($\ell=500$)  &     $54.00 \pm 0.31$     &      $27.19 \pm 0.38$  &     $14.79 ^\dagger $ \\ 
HQD ResNeXt-50  ($\ell=1000$)  &     $54.02 \pm 0.31 $     &      $26.82 \pm 0.35$  &     $14.79 ^\dagger $ \\ 
\bottomrule

\multicolumn{4}{l}{ $\dagger$ indicates $9$ more parameters in the HQD variational circuit.}\\

\multicolumn{4}{l}{ We color each cell as \colorbox{red!50}{best} and \colorbox{yellow!40}{second best} and \textbf{best} means better than the classical approach.}
\end{tabular}
}
\label{CIFAR100-2}
\end{table*}
Furthermore, it's important to highlight that while HQD Xception shows improvement on the CIFAR-10 dataset, its performance on CIFAR-100 sees negligible enhancement from the HQD module, regardless of the $\ell$ value except $\ell= 100$ for top-5 error in Table. \ref{CIFAR100}. CIFAR-100 has more categories and higher intra-category variability than CIFAR-10, which requires the model to be able to extract more complex and detailed features. Although the Xception model performs well in extracting and managing high-level features, efficiently utilizing parameters through depth-separable convolution, and conducting fine spatial and channel analysis of features, the effect of HQD may be offset by the complexity of the task itself. Conversely, it is observed that Xception exhibited the lowest top-1 and top-5 errors across all tested models, yet its performance remained unaffected by HQD integration. A potential reason is that the model is trying to adapt to more complex data structures and category subdivisions, and generalization capabilities are sacrificed by overreliance on specific pooling strategies. In this case, the model may have learned too specific feature expressions on the training set, and these feature expressions may not be able to generalize to the test set effectively. A possible solution is to carry out more mixing ratios $(\ell)$ of quantum downsampling and maximum pooling to adapt to the high complexity of CIFAR-100 for the Xception model.

\begin{table*}[ht]
\centering
\caption{Experimental Results on CIFAR-100 for Attention-56 and SE-ResNet-50.}
\setlength{\tabcolsep}{4pt} 
%\fontsize{10pt}{10pt}\selectfont
\resizebox{0.9\textwidth}{!}{
\begin{tabular}{lccc}

\toprule
               Method     & Top-1 err. ($\%$) $(\downarrow)$ & Top-5 err. ($\%$)$(\downarrow)$ & Parameters (M) $(\downarrow)$\\  
                    \midrule
 Attention-56 \cite{Attention}  &   $56.37$  
    &      $30.47$     &     $55.70$
     \\ 
\midrule  
HQD Attention-56 ($\ell=50$)  &     $ 57.83 \pm 0.17 $     &      $31.26 \pm 0.15 $  &     $55.70 ^\dagger $ \\ 
HQD Attention-56 ($\ell=100$)  &     $\cellcolor{red!50} 55.69 \pm 0.09 $     &      $\cellcolor{red!50}29.67 \pm 0.12 $  &     $55.70 ^\dagger $ \\ 
HQD Attention-56 ($\ell=500$)  &     $56.58 \pm 0.13 $     &      $\cellcolor{yellow!40}30.28 \pm 0.15 $  &     $55.70 ^\dagger $ \\ 
HQD Attention-56 ($\ell=1000$)  &     $57.66 \pm 0.11 $     &      $31.25 \pm 0.09$  &     $55.70 ^\dagger $ \\ 
\midrule 
SE-ResNet-50 \cite{Hu2017SqueezeandExcitationN} &     $48.06$     &      $21.45$     &     $28.78 $\\ 
\midrule  
HQD SE-ResNet-50 ($\ell=50$)  &     $47.64 \pm 0.07 $     &      $21.47 \pm 0.17$  &     $28.78 ^\dagger $ \\ 
HQD SE-ResNet-50 ($\ell=100$)  &     $\cellcolor{orange!40}47.48 \pm 0.08 $     &      $\cellcolor{red!50}20.79 \pm 0.06 $  &     $28.78 ^\dagger $ \\ 
HQD SE-ResNet-50 ($\ell=500$)  &     $\cellcolor{red!50}47.41 \pm 0.13 $     &      $\cellcolor{yellow!40}21.27 \pm 0.09 $  &     $28.78 ^\dagger $ \\ 
HQD SE-ResNet-50 ($\ell=1000$)  &     $\cellcolor{yellow!40}47.47 \pm 0.15$     &      $\cellcolor{orange!40}21.38 \pm 0.14 $  &     $28.78 ^\dagger  $ \\ 
\bottomrule
\multicolumn{4}{l}{$\dagger$ indicates $9$ more parameters in the HQD variational circuit, and \colorbox{red!50}{best}, \colorbox{yellow!40}{second best} and \colorbox{orange!40}{third best}.}
\end{tabular}
}

\label{CIFAR100-3}
\end{table*}

\begin{table*}[ht]
\centering
\caption{Experimental Results on CIFAR-100 for Xception and SqueezeNet.}
\setlength{\tabcolsep}{4pt} 
\resizebox{0.9\textwidth}{!}{
%\fontsize{10pt}{10pt}\selectfont
\begin{tabular}{lccc}

\toprule
              Method     & Top-1 err. ($\%$) $(\downarrow)$ & Top-5 err. ($\%$)$(\downarrow)$ & Parameters (M) $(\downarrow)$\\ 
                    \midrule
Xception \cite{Chollet2016XceptionDL}  &   $41.53$ 
    &      $17.01$  &     $21.01$ \\
\midrule  
HQD Xception ($\ell=50$)  &     $42.43 \pm 0.09 $     &      $17.93 \pm 0.13$  &     $21.01 ^\dagger $ \\ 
HQD Xception ($\ell=100$)  &     $41.69 \pm 0.03$     &      $\cellcolor{red!50}16.97 \pm 0.15 $  &     $21.01 ^\dagger $ \\ 
HQD Xception  ($\ell=500$)  &     $42.19 \pm 0.08$     &      $17.61 \pm 0.11$  &     $21.01 ^\dagger $ \\ 
HQD Xception ($\ell=1000$)  &     $42.47 \pm 0.12 $     &      $17.22 \pm 0.16 $  &     $21.01 ^\dagger $ \\ 
\midrule
SqueezeNet \cite{Iandola2016SqueezeNetAA} &     $47.27$     &      $20.45$     &     $0.78 $\\ 
\midrule  
HQD SqueezeNet ($\ell=50$)  &     $\cellcolor{red!50}44.75 \pm 0.29 $     &      $\cellcolor{yellow!40}18.69 \pm 0.21 $  &     $0.78 ^\dagger $ \\  
HQD SqueezeNet ($\ell=100$)  &     $\cellcolor{yellow!40}45.71 \pm 0.23 $     &      $\cellcolor{red!50}18.55 \pm 0.27$  &     $0.78 ^\dagger $ \\  
HQD SqueezeNet ($\ell=500$)  &     $\cellcolor{orange!40}45.99 \pm 0.30 $     &      $\cellcolor{orange!40}19.30 \pm 0.33 $  &     $0.78 ^\dagger  $ \\  
HQD SqueezeNet ($\ell=1000$)  &     $46.02 \pm 0.36$     &      $19.46 \pm 0.29$  &     $0.78 ^\dagger $ \\  
\bottomrule
\multicolumn{4}{l}{ $\dagger$ indicates $9$ more parameters in the HQD variational circuit, and \colorbox{red!50}{best}, \colorbox{yellow!40}{second best} and \colorbox{orange!40}{third best}.}
\end{tabular}
}

\label{CIFAR100}
\end{table*}

Additionally, it was noted that the proposed HQD module exhibited resilience to noise across various experiments, with test results varying within a narrow margin of $\pm 0.4$. One explanation for this is that the HQD model treats quantum noise as part of the deep neural network and adapts to these noises during training. Through gradient descent and optimization one at a time, the parameters of the deep neural network and the parameters in the variational quantum circuit in the HQD module are optimized in this process. This experimental result also proves the robustness of our proposed HQD module.

\section{Conclusion}
In this study, we proposed an advanced hybrid quantum-classical paradigm, the HQD module, which was predicated upon an innovative mixed pooling methodology. Our pioneering experiments provide empirical evidence that the HQD module can yield a discernible quantum advantage in addressing traditional challenges in computer vision tasks with noise-resilient capabilities. During downsampling, quantum computing can capture more nonlinear features while retaining as much detail as possible. It is worth mentioning that the HQD module can be seamlessly integrated into any SOTA deep learning model. Meanwhile, this novel strategy has the potential to catalyze the development of superior algorithms for a spectrum of applications in computer vision and beyond. We envision that the HQD module will play a vital and meaningful role in image segmentation and edge detection tasks and pave a crucial pathway for the advancement of quantum artificial intelligence (QAI).

\label{conclusion}

\clearpage  % TODO REVIEW/FINAL: This \clearpage needs to be removed from both review and camera-ready versions.

% ---- Bibliography ----
%
% BibTeX users should specify bibliography style 'splncs04'.
% References will then be sorted and formatted in the correct style.
%

\bibliographystyle{plainnat}
\bibliography{main}
\clearpage

\appendix
\section*{Appendix}

\begin{algorithm*}
	%\textsl{}\setstretch{1.8}
	\renewcommand{\algorithmicrequire}{\textbf{Input:}}
	\renewcommand{\algorithmicensure}{\textbf{Output:}}
	\caption{HQD Module}
        \textbf{Input}: Kernel size: $K$, Padding: $p$, Stride: $s$, training $epoch$\\
        \textbf{Input}: Width and height of the pooling window: $a \times b$, $a$ and $b$ are the pixel index in kernel window $ (K-1)\times (K-1) $, hybrid parameter: $\ell$ \\
        \textbf{Initialize}: Quantum state: $\rho = |0\rangle$ 
        \begin{algorithmic}[1]
        \FOR{$i=1$ to $epoch$}
        \FOR{$l_0=1$ to $\ell$}
            \IF{$l_0 \leq \ell - 1$}
            \STATE \textbf{Max pooling:}\\
            \STATE Evaluate: $H_{out}$ and $W_{out}$
            \STATE Compute: \\
            $\tilde{I_M}(i,j) =$ \\
            $\max_{a=0}^{K-1} \max_{b=0}^{K-1} I(s \cdot i + a - p, s \cdot j + b - p)$
            \ENDIF
            \STATE \textbf{Quantum Downsampling:}
            \STATE Evaluate: 
    ${\rm P}_{r} \left( l \right) = \langle \rho  {|\mathcal{O}_l^\dag  \mathcal{O}_l \left| {\rho } \right.} \rangle $
  
        \ENDFOR
        \ENDFOR
        \RETURN Feature map $\tilde{I_M}: H_{out} \times W_{out}$.
        \end{algorithmic}
        \label{ALGO-1}
\end{algorithm*}

\section{Limitations}

\textbf{Parameters.} Due to the quantum variational circuits, there are $9$ more parameters to be optimized in training compared to the classical max pooling method in the various deep learning architectures, which means model training and inference will take longer. However, in Fig. \ref{macparamcifar10}, we can see that the introduction of $9$ additional parameters does not significantly increase the computational complexity in the training process. However, increasing parameters means the model requires more memory, which is particularly problematic on resource-constrained devices. Although the additional parameters of quantum downsampling introduce some challenges, better computing resources can minimize the impact of these limitations while improving model performance.

\textbf{Quantum Hardware.} Considering the $3 \times 3$ pooling operation, it is noteworthy that merely $9$ qubits are requisite for a quantum computer to fulfill the entire procedure. However, the limitation is twofold: firstly, the current state of quantum hardware, characterized by its qubit fidelity and coherence time, poses challenges to the stability and reliability of executing operations over even a modest number of qubits \cite{de2021materials}. Secondly, the integration of quantum processes into classical deep learning workflows requires quantum-classical interface mechanisms that can maintain the integrity of the computational process. These considerations underscore the necessity for advancements in quantum hardware and algorithmic strategies, particularly in the context of tasks like pooling operations that are ubiquitous in deep learning architectures. 

\textbf{Quantum Noise.} Although quantum noise simulation experiments can provide insights into the theoretical resistance of algorithms to noise, these simulations often have significant gaps with reality. Furthermore, quantum computers also face the challenge of quantum state preparation and readout when processing such large-scale datasets. Likewise, readout errors during the quantum measurement process may also seriously affect the accuracy of the classification results \cite{preskill2018quantum}. Current quantum technologies are still limited in precise control and measurement, which is particularly problematic when processing large amounts of data. While simulation experiments provide important insights into how quantum algorithms perform in the face of noise, applying these algorithms to actual quantum hardware and processing large-scale datasets requires a lot of effort. 

\section{CIFAR-10 dataset}
\begin{figure}[ht]

\begin{center}

\centerline{\includegraphics[width=0.91\textwidth]{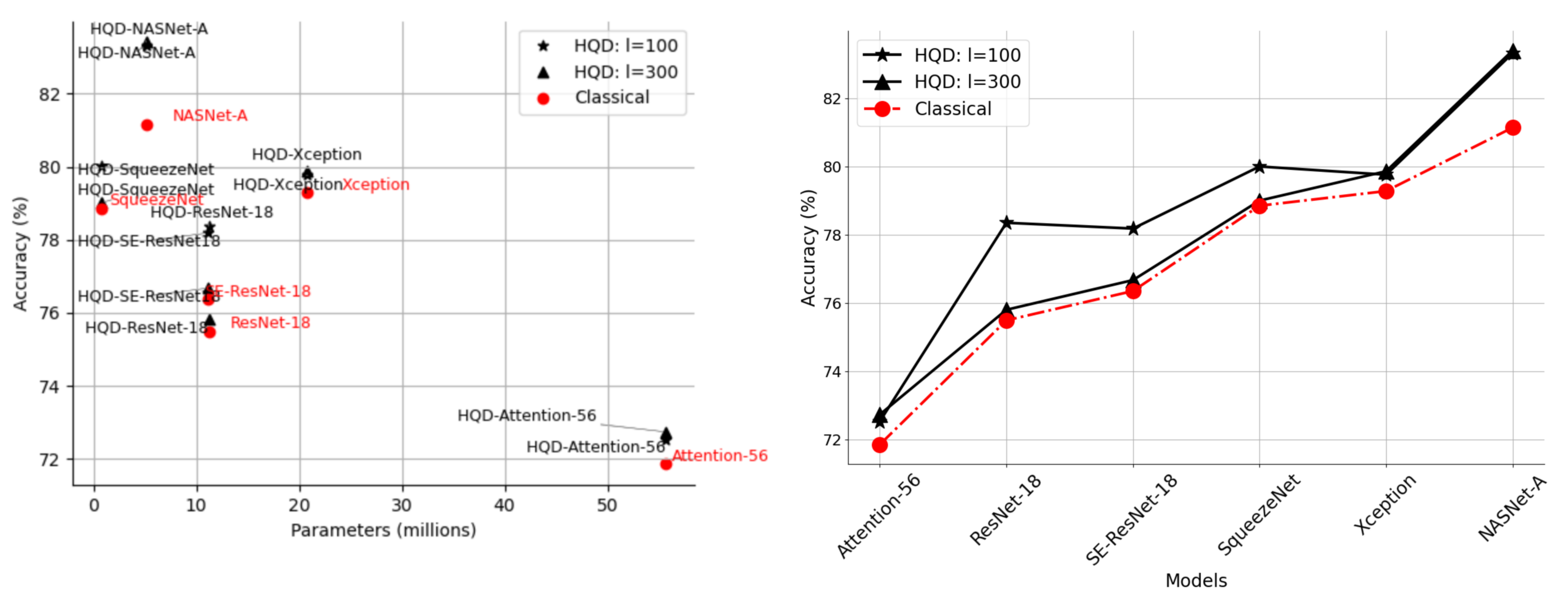}}
\caption{Accuracy versus parameters (left) and accuracy versus different models (right) on CIFAR-10 dataset. Red circles indicate the classical results, and the black shapes stand for our proposed models.}
\label{macparamcifar10}

\end{center}
\end{figure}

 By correlating the data on the number of parameters depicted in Fig. \ref{macparamcifar10} with the more illustrative trend line chart, it becomes apparent that an increase in model parameters theoretically enhances the model's representational capacity. However, this augmentation simultaneously elevates the likelihood of overfitting. For high-parameter models, the incremental benefits from the HQD module may be limited. These models already capture extensive feature information, making the performance gains from further optimizations less significant compared to models with fewer parameters. Conversely, varying $\ell$ values impact models differently. A higher $\ell$ value indicates reduced involvement of the HQD module.
\section{The $2 \times 2$ HQD Module and Application}
\begin{figure*}[h]

\begin{center}
\centerline{\includegraphics[width=1\textwidth]{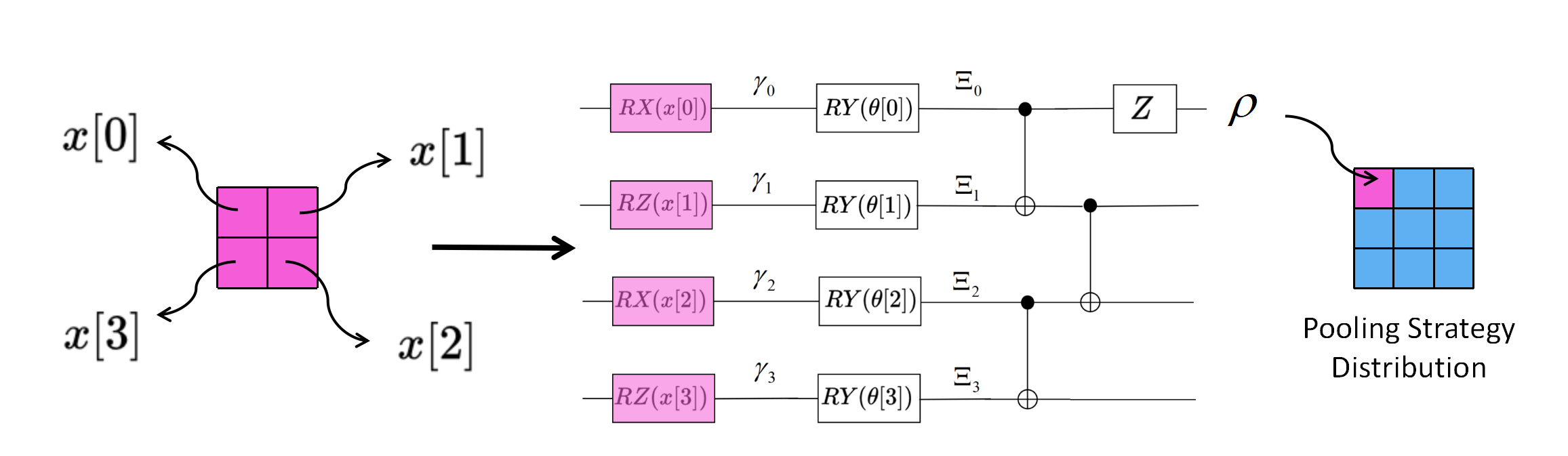}}
\caption{Structure of $2 \times 2$ quantum downsampling: it shows the $2 \times 2$ pooling window, extracts $4$ eigenvalues, and puts them into the quantum variational circuit to obtain an output. \textbf{Pink} color indicates the use of \textbf{quantum downsampling}. \textbf{Pink} color in the quantum variational circuit $x[0]-x[3]$ are the value from the input image, and $\theta$ is the parameters to be optimized. \textbf{Blue} color indicates the use of \textbf{classical max pooling}. The pooling strategy distribution is determined by $\ell$.}
\label{quantumpooling2X2}

\end{center}

\end{figure*}

In Fig. \ref{quantumpooling2X2}, the quantum variational circuit for downsampling with a $2 \times 2$ kernel is presented. A design similar to the $3 \times 3$ quantum variational circuit connects the $RX$ gate to the $RY$ gate and the $RZ$ gate to the $RY$ gate. Then, the CNOT gate is used to create entanglement between them, and finally, the Pauli-Z gate is used to measure the result. Due to the space limitation, we implement the $2 \times 2$ quantum downsampling module in the VGG structure in the following section.

\clearpage

\begin{table*}[ht]
\centering
\caption{Experimental Results on CIFAR-100 for VGG-13, VGG-16 and VGG -19.}
\setlength{\tabcolsep}{4pt} 
%\fontsize{10pt}{10pt}\selectfont
\resizebox{0.95\textwidth}{!}{
\begin{tabular}{lccc}

\toprule
             Method     & Top-1 err. ($\%$) $(\downarrow)$ & Top-5 err. ($\%$)$(\downarrow)$ & Parameters (M) $(\downarrow)$\\   
                    \midrule
VGG-13 \cite{Simonyan2014VeryDCvgg16}  &   $40.86$ 
    &      $16.31$  &     $28.15$ \\
\midrule  
HQD VGG-13 ($\ell=50$)  &     $41.64 \pm 0.12 $     &      $17.54 \pm 0.13 $  &     $28.15 ^ \ddagger $ \\ 
HQD VGG-13 ($\ell=100$)  &     $41.27 \pm 0.35 $     &    $16.45\pm 0.29 $  &     $28.15 ^ \ddagger$ \\ 
HQD VGG-13 ($\ell=200$)  &   $41.01 \pm 0.45 $     &      $16.43 \pm 0.01$  &     $28.15 ^ \ddagger$ \\ 
HQD VGG-13 ($\ell=300$)  &  \cellcolor{red!50}$40.71 \pm 0.22$     &   \cellcolor{yellow!40}$16.07 \pm 0.18$  &     $28.15 ^ \ddagger$ \\ 
HQD VGG-13 ($\ell=400$)  &    $41.13 \pm 0.24$     &      \cellcolor{red!50}$16.01 \pm 0.11 $  &     $28.15 ^ \ddagger$ \\ 
HQD VGG-13  ($\ell=500$)  &     $41.24 \pm 0.31 $     &      $16.36 \pm 0.22 $  &     $28.15 ^ \ddagger $ \\ 
HQD VGG-13 ($\ell=1000$)  &     $41.61 \pm 0.04 $     &    \cellcolor{orange!40}$16.11 \pm 0.09$  &     $28.15 ^ \ddagger $ \\ 
\midrule 
VGG-16 \cite{Simonyan2014VeryDCvgg16}  &   $41.25$ 
    &      $16.36$  &     $33.65$ \\
\midrule  
HQD VGG-16 ($\ell=50$)  &     $41.17 \pm 0.11$     &      $17.60 \pm 0.02$  &     $33.65 ^ \ddagger $ \\ 
HQD VGG-16 ($\ell=100$)  &     $40.94 \pm 0.24 $     &      \cellcolor{orange!40}$16.14 \pm 0.27 $  &     $33.65 ^ \ddagger$ \\ 
HQD VGG-16 ($\ell=200$)  &  \cellcolor{brown!40}$40.84 \pm 0.21$     &   \cellcolor{yellow!40}$16.09 \pm 0.03$  &     $33.65 ^ \ddagger$ \\ 
HQD VGG-16 ($\ell=300$)  &   \cellcolor{red!50}$40.50 \pm 0.33 $     &    \cellcolor{red!50}$15.92 \pm 0.32 $  &     $33.65 ^ \ddagger$ \\ 
HQD VGG-16 ($\ell=400$)  &     $41.12 \pm 0.23 $     &      $16.52 \pm 0.19 $  &     $33.65 ^ \ddagger$ \\ 
HQD VGG-16 ($\ell=500$)  &   \cellcolor{yellow!40}$40.60 \pm 0.17 $     &     $16.28 \pm 0.22 $  &     $33.65 ^ \ddagger $ \\ 
HQD VGG-16 ($\ell=1000$)  &    \cellcolor{orange!40}$40.67 \pm 0.15 $     &   \cellcolor{brown!40}$16.24 \pm 0.18 $  &     $33.65 ^ \ddagger $ \\ 
\midrule 
VGG-19 \cite{Simonyan2014VeryDCvgg16}  &   $42.36 $ 
    &      $18.33 $  &     $38.96$ \\
\midrule  
HQD VGG-19 ($\ell=50$)  &   \cellcolor{yellow!40}$41.40 \pm 0.12 $     &      $17.76 \pm 0.07$  &     $38.96 ^ \ddagger $ \\ 
HQD VGG-19 ($\ell=100$)  &     $41.57 \pm 0.26 $     &      \cellcolor{red!50}$17.10 \pm 0.65 $  &     $38.96 ^ \ddagger$ \\ 
HQD VGG-19 ($\ell=200$)  &  \cellcolor{red!50}$41.37 \pm 0.19 $     &   $17.55 \pm 0.19$  &     $38.96 ^ \ddagger$ \\ 
HQD VGG-19 ($\ell=300$)  &    \cellcolor{orange!40}$41.48 \pm 0.46 $     &  \cellcolor{orange!40}$17.35 \pm 0.03$  &     $38.96 ^ \ddagger$ \\ 
HQD VGG-19 ($\ell=400$)  &   \cellcolor{brown!40}  $41.53 \pm 0.31 $     &      $17.74 \pm 0.27 $  &     $38.96 ^ \ddagger$ \\ 
HQD VGG-19 ($\ell=500$)  &     $41.81 \pm 0.25 $     &  \cellcolor{brown!40}$17.44 \pm 0. 27 $  &     $38.96 ^ \ddagger$ \\ 
HQD VGG-19 ($\ell=1000$)  &     $41.85 \pm 0.14 $     &    \cellcolor{yellow!40}$17.20 \pm 0.11 $  &     $38.96 ^ \ddagger $ \\ 

\bottomrule

\multicolumn{4}{l}{\textbf{Best} means better than the classical approach.}\\
\multicolumn{4}{l}{ We color each cell as \colorbox{red!50}{best}, \colorbox{yellow!40}{second best}, \colorbox{orange!40}{third best} and \colorbox{brown!40}{fourth best}.}\\
\multicolumn{4}{l}{ $\ddagger$ indicates $4$ more parameters in the HQD variational circuit for kernel size $2 \times 2$.}
\end{tabular}
}

\label{CIFAR100vgg2}
\end{table*}

Table. \ref{CIFAR100vgg2} illustrates that the HQD module yields the most significant enhancement in performance for the VGG-19 architecture, followed by VGG-16, while the improvement observed for VGG-13 is minimal. Notably, within the context of Top-1 error metrics, only a singular configuration manages ($\ell = 300$) to improve performance. One possible reason is that the VGG-19 and VGG-16 models have more convolutional layers and parameters than VGG-13 as shown in Fig. \ref{VGG}, which means they are able to benefit from more complex feature extraction and finer-grained feature representation. The HQD module can capture and deliver important feature information more efficiently through its design, which may be more evident in deeper and more complex networks. At the same time, the HQD module is designed to capture and retain important feature information while reducing information loss. In deeper networks, the optimization of information transfer and feature representation may be more critical, so the HQD module improves VGG-19 and VGG-16 more significantly.

From Table. \ref{CIFAR100}, we can also see that classical VGG-13 has the best performance among the classic models. This is because deeper models such as VGG-19 and VGG-16 are more likely to overfit in this case. The HQD module provides an effective form of regularization for these models by improving feature selection and transfer, thereby mitigating overfitting problems while maintaining model capacity. Overfitting is not a major problem for shallower models like VGG-13, so the HQD module's improvement is not so obvious.

\clearpage

\section{Plots of Table. 2, Table. 3 and Table. 4 in the Main Paper}

\begin{figure}[ht]
\begin{center}
\centerline{\includegraphics[width=1\textwidth]{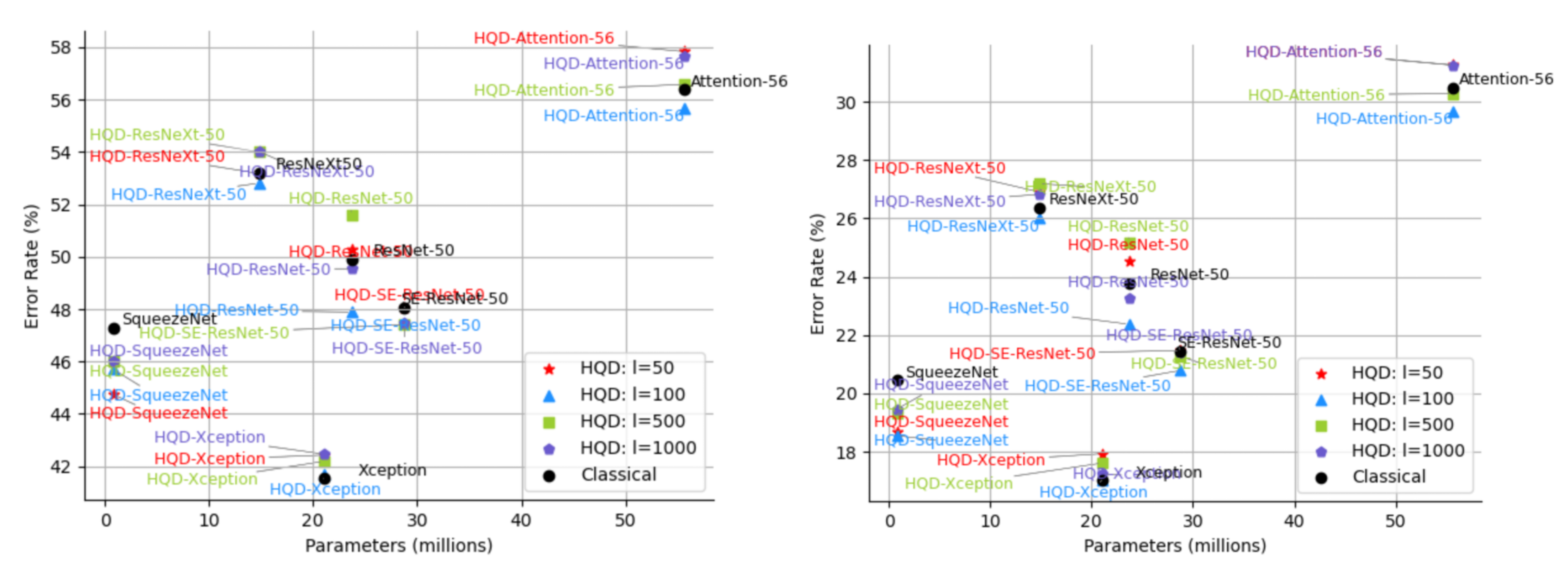}}
\caption{Top-1 error (left) and Top-5 error (right) versus parameters on CIFAR-100 dataset.}
\label{scatter15}
\end{center}

\end{figure}

\begin{figure}[ht]
\begin{center}
\centerline{\includegraphics[width=1\textwidth]{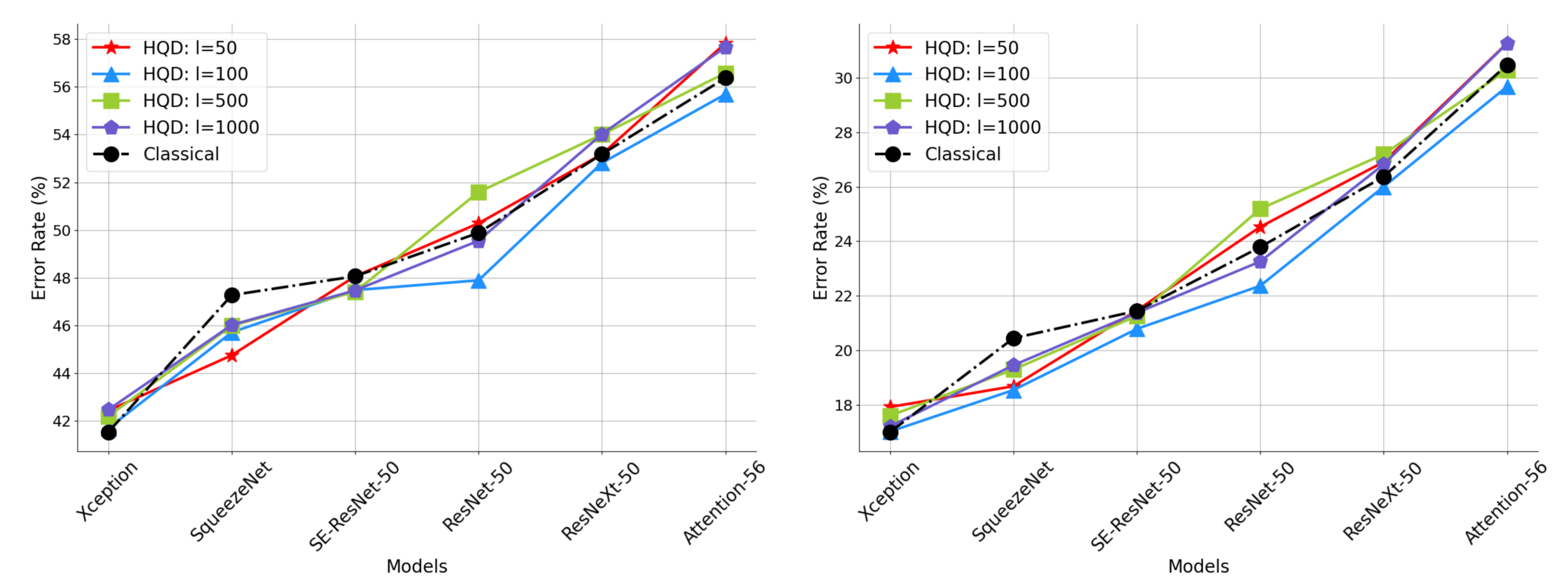}}
\caption{Top-1 error (left) and Top-5 error (right) versus different models on CIFAR-100 dataset.}
\label{paramacc}
\end{center}
\end{figure}

Fig. \ref{scatter15} and \ref{paramacc} reveal varying degrees of performance enhancements across different models attributable to the HQD module. Specifically, the figures highlight an improvement in SqueezeNet's performance across different $\ell$ values. Regarding the Top-1 error metric, the application of the HQD module, with its varying $\ell$ values, introduces the most significant performance variability in ResNet-50.

\clearpage

\section{VGG and HQD VGG Structures}
\label{secvgg}
\begin{figure*}

\begin{center}
\centerline{\includegraphics[width=0.75\textwidth]{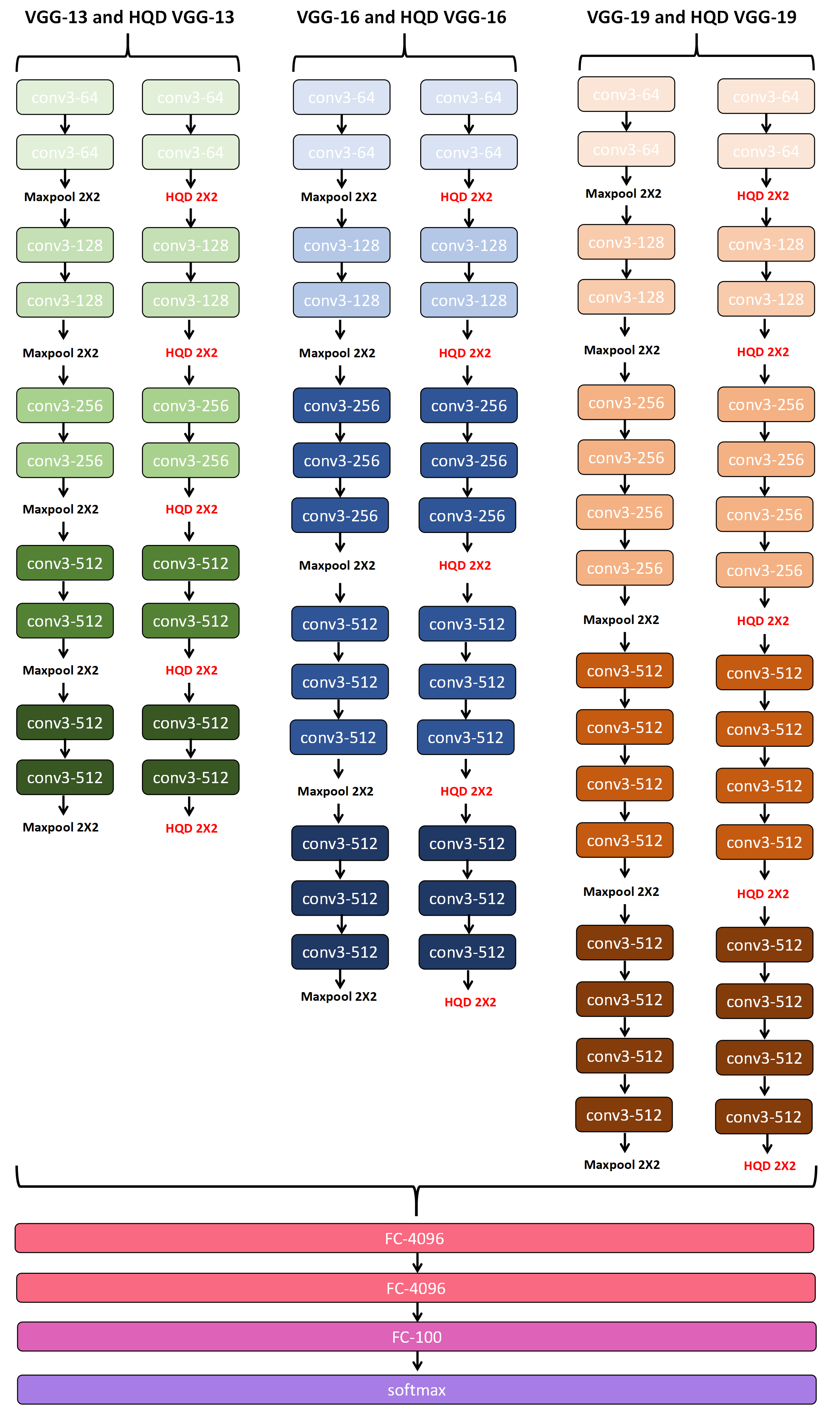}}
\caption{Structure of VGG (Left) and HQD VGG (Right).}
\label{VGG}
\end{center}

\end{figure*}

In Fig. \ref{VGG}, we showed three kinds of VGG structures, VGG-13, VGG-16, and VGG-19 respectively. For each kind of VGG structure, HQD networks are shown on the right hand. More specifically, the $2\times2$ max pooling layer in the original VGG structure was replaced by the HQD module with $4$ qubits.

\clearpage

\section{SqueezeNet and HQD SqueezeNet Structures}

\begin{figure*}[h]

\begin{center}
\centerline{\includegraphics[width=0.7\textwidth]{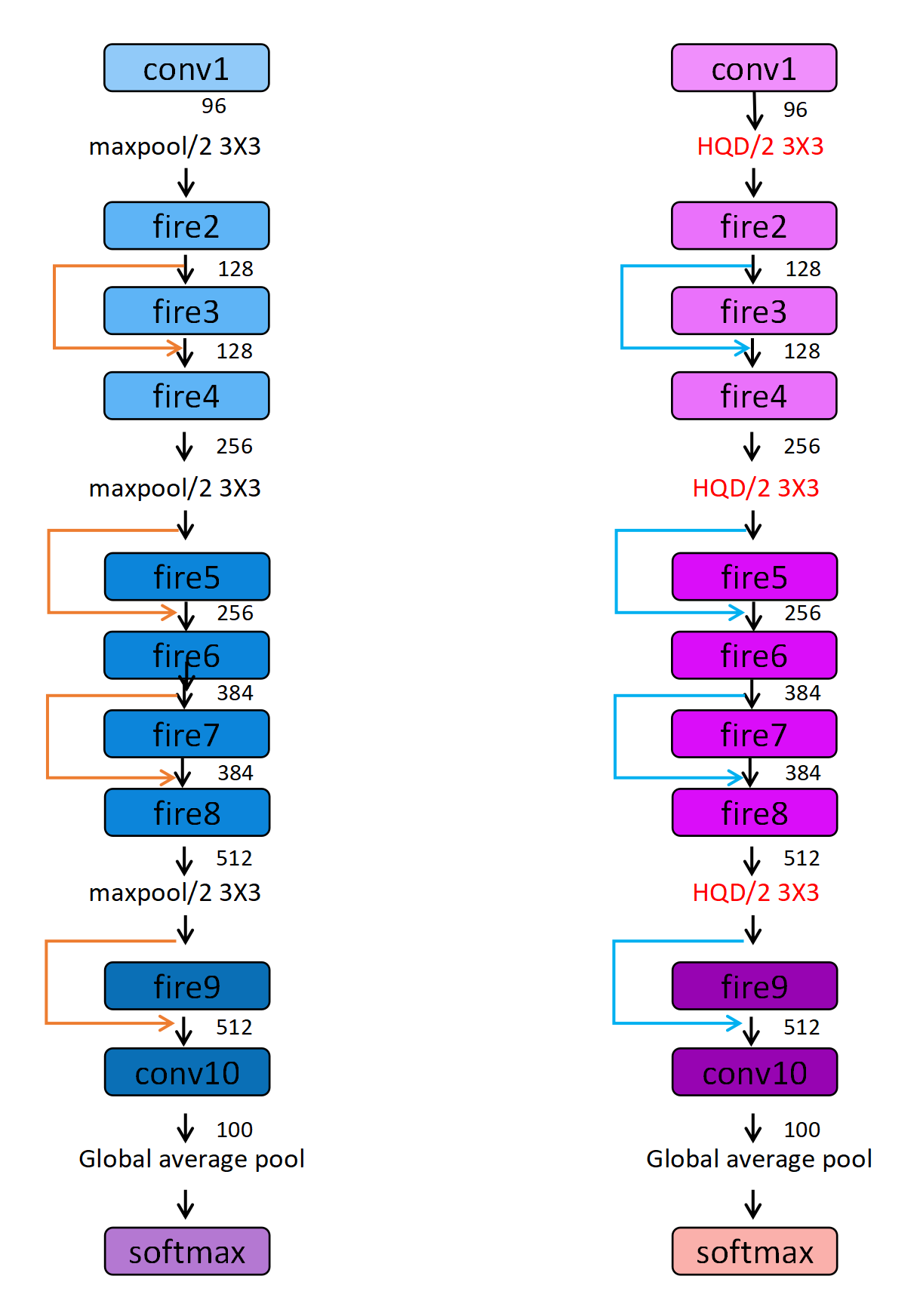}}
\caption{Structure of SqueezeNet (Left) and HQD SqueezeNet (Right).}
\label{squeezenet}
\end{center}
\end{figure*}

In Fig. \ref{squeezenet}, we showed the structure of the classical SqueezeNet model on the left hand \cite{Iandola2016SqueezeNetAA} and the HQD SqueezeNet model on the right hand.

\clearpage

\section{Attention Networks and HQD Attention Networks Structures}

\begin{table}[ht]
\centering
\caption{Attention-56 and Attention-92 \cite{Attention} Structure}
\resizebox{0.6\textwidth}{!}{
\begin{tabular}{c|c|cc}
\hline
Layer & Output Size & \multicolumn{1}{c|}{Attention-56} & Attention-92 \\ \hline
Conv1 & $112 \times 112 $ & \multicolumn{2}{c}{$7 \times 7, 64,\text{stride}2$}   \\ \hline
  Max pooling & $56 \times 56$ & \multicolumn{2}{c}{$3 \times 3, \text{stride}2$}    \\ \hline
  
Residual Unit  & $56 \times 56$ & \multicolumn{2}{c}{$\begin{pmatrix}
 1 \times 1 & 64 \\
 3 \times 3 &  64 \\
 1 \times 1 & 256 
\end{pmatrix} \times 1$}    \\ \hline

Attention Module & $ 56 \times 56 $ & \multicolumn{1}{c|}{Attention $\times 1$} & Attention $\times 1$ \\ \hline

Residual Unit  & $28 \times 28$ & \multicolumn{2}{c}{$\begin{pmatrix}
 1 \times 1 & 128 \\
 3 \times 3 & 128 \\
 1 \times 1 & 512 
\end{pmatrix} \times 1$}    \\ \hline

Attention Module & $ 28 \times 28 $ & \multicolumn{1}{c|}{Attention $\times 1$} & Attention $\times 2$ \\ \hline

Residual Unit  & $14 \times 14$ & \multicolumn{2}{c}{$\begin{pmatrix}
 1 \times 1 & 256 \\
 3 \times 3 & 256 \\
 1 \times 1 & 1024 
\end{pmatrix} \times 1$}    \\ \hline

Attention Module & $ 14 \times 14 $ & \multicolumn{1}{c|}{Attention $\times 1$} & Attention $\times 3$ \\ \hline

Residual Unit  & $7 \times 7$ & \multicolumn{2}{c}{$\begin{pmatrix}
 1 \times 1 & 512\\
 3 \times 3 & 512 \\
 1 \times 1 & 2048
\end{pmatrix} \times 3$}    \\ \hline

Average pooling  & $1 \times 1$ & \multicolumn{2}{c}{$7 \times 7, \text{stride} 1$}    \\ \hline
Softmax  &  \multicolumn{3}{c}{$100$}    \\ \hline

\end{tabular}
}
\end{table}

\begin{table}[ht]
\centering
\caption{HQD Attention-56 and HQD Attention-92 Structure}
\resizebox{0.6\textwidth}{!}{
\begin{tabular}{c|c|cc}
\hline
Layer & Output Size & \multicolumn{1}{c|}{Attention-56} & Attention-92 \\ \hline
Conv1 & $112 \times 112 $ & \multicolumn{2}{c}{$7 \times 7, 64,\text{stride}2$}   \\ \hline
  HQD Module & $56 \times 56$ & \multicolumn{2}{c}{$3 \times 3, \text{stride}2$}    \\ \hline
  
Residual Unit  & $56 \times 56$ & \multicolumn{2}{c}{$\begin{pmatrix}
 1 \times 1 & 64 \\
 3 \times 3 &  64 \\
 1 \times 1 & 256 
\end{pmatrix} \times 1$}    \\ \hline

Attention Module & $ 56 \times 56 $ & \multicolumn{1}{c|}{Attention $\times 1$} & Attention $\times 1$ \\ \hline

Residual Unit  & $28 \times 28$ & \multicolumn{2}{c}{$\begin{pmatrix}
 1 \times 1 & 128 \\
 3 \times 3 & 128 \\
 1 \times 1 & 512 
\end{pmatrix} \times 1$}    \\ \hline

Attention Module & $ 28 \times 28 $ & \multicolumn{1}{c|}{Attention $\times 1$} & Attention $\times 2$ \\ \hline

Residual Unit  & $14 \times 14$ & \multicolumn{2}{c}{$\begin{pmatrix}
 1 \times 1 & 256 \\
 3 \times 3 & 256 \\
 1 \times 1 & 1024 
\end{pmatrix} \times 1$}    \\ \hline

Attention Module & $ 14 \times 14 $ & \multicolumn{1}{c|}{Attention $\times 1$} & Attention $\times 3$ \\ \hline

Residual Unit  & $7 \times 7$ & \multicolumn{2}{c}{$\begin{pmatrix}
 1 \times 1 & 512\\
 3 \times 3 & 512 \\
 1 \times 1 & 2048
\end{pmatrix} \times 3$}    \\ \hline

Average pooling  & $1 \times 1$ & \multicolumn{2}{c}{$7 \times 7, \text{stride} 1$}    \\ \hline
Softmax  &  \multicolumn{3}{c}{$100$}    \\ \hline

\end{tabular}
}
\end{table}

\clearpage

\section{SE-ResNet, ResNet, HQD SE-ResNet and HQD ResNet Structures}

\label{appendixres18}
% You can have as much text here as you want. The main body must be at most $8$ pages long.
% For the final version, one more page can be added.
% If you want, you can use an appendix like this one.  

% The $\mathtt{\backslash onecolumn}$ command above can be kept in place if you prefer a one-column appendix, or can be removed if you prefer a two-column appendix.  Apart from this possible change, the style (font size, spacing, margins, page numbering, etc.) should be kept the same as the main body.
% %%%%%%%%%%%%%%%%%%%%%%%%%%%%%%%%%%%%%%%%%%%%%%%%%%%%%%%%%%%%%%%%%%%%%%%%%%%%%%%
% %%%%%%%%%%%%%%%%%%%%%%%%%%%%%%%%%%%%%%%%%%%%%%%%%%%%%%%%%%%%%%%%%%%%%%%%%%%%%%%

\begin{table*}[htbp]
\begin{center}
\caption{ResNet-18 \cite{Resnet} and HQD ResNet-18 Structure}
\resizebox{1\textwidth}{!}{
\begin{tabular}{cccc}
\hline
Layer Name                & Output Size       & ResNet-18 & HQD ResNet-18 \\ \hline
conv1                     &       $112 \times 112 \times 64 $            &     $ 7 \times 7$, $64$, stride 2      &      $ 7 \times 7$, $64$, stride 2        \\ \hline
 &     &    $3 \times 3$ max pool, stride 2       &     $3 \times 3$ HQD module          \\ \cline{3-4} 
{conv2\_x} &   $56 \times 56 \times 64 $       &       \rule{0pt}{5ex} $\begin{bmatrix}
 3 \times 3, &64 \\
  3 \times 3, &64
\end{bmatrix} \times 2 \rule[-4ex]{0pt}{8ex}$     &        $\begin{bmatrix}
 3 \times 3, &64 \\
  3 \times 3, &64 
\end{bmatrix} \times 2 $       \\ \hline
conv3\_x                  &       $28 \times 28 \times 128 $            &      \rule{0pt}{4ex} $\begin{bmatrix}
 3 \times 3, &128 \\
  3 \times 3, &128
\end{bmatrix} \times 2 \rule[-4ex]{0pt}{8ex}$      &      \rule{0pt}{5ex} $\begin{bmatrix}
 3 \times 3, &128 \\
  3 \times 3, &128
\end{bmatrix} \times 2 \rule[-3ex]{0pt}{6ex}$        \\ \hline
conv4\_x                  &           $14\times 14 \times 256 $        &    \rule{0pt}{4ex} $\begin{bmatrix}
 3 \times 3, &256 \\
  3 \times 3, &256
\end{bmatrix} \times 2 \rule[-3ex]{0pt}{6ex}$       &         \rule{0pt}{4ex} $\begin{bmatrix}
 3 \times 3, &256 \\
  3 \times 3, &256 
\end{bmatrix} \times 2 \rule[-3ex]{0pt}{6ex}$     \\ \hline
conv5\_x                  &         $7 \times 7 \times 512 $          &     \rule{0pt}{4ex} $\begin{bmatrix}
 3 \times 3, &512 \\
  3 \times 3, &512
\end{bmatrix} \times 2 \rule[-4ex]{0pt}{6ex}$      &        \rule{0pt}{5ex} $\begin{bmatrix}
 3 \times 3, &512 \\
  3 \times 3, &512 
\end{bmatrix} \times 2 \rule[-3ex]{0pt}{6ex}$      \\ \hline
average pool              &          $1 \times 1 \times 64 $         &     $7 \times 7$  average pool    &         $7 \times 7$  average pool      \\ \hline
fully connected           &           $10$        &     $512 \times 10$ &          $512 \times 10$    \\ \hline
softmax                   &            $10$       &           &              
\end{tabular}

}
\end{center}
\end{table*}

\begin{table*}[htbp]
\begin{center}
\caption{SE-ResNet-18 \cite{Hu2017SqueezeandExcitationN} and HQD SE-ResNet-18 Structure}
\resizebox{1\textwidth}{!}{
\begin{tabular}{cccc}
\hline
Layer Name                & Output Size       & SE-ResNet-18 & HQD SE-ResNet-18 \\ \hline
conv1                     &       $112 \times 112 \times 64 $            &     $ 7 \times 7$, $64$, stride 2      &      $ 7 \times 7$, $64$, stride 2        \\ \hline
 &     &    $3 \times 3$ max pool, stride 2       &     $3 \times 3$ HQD module          \\ \cline{3-4} 
{conv2\_x} &   $56 \times 56 \times 64 $       &       \rule{0pt}{5ex} $\begin{bmatrix}
 3 \times 3, &64 \\
  3 \times 3, &64 \\
  fc, & [2,64]
\end{bmatrix} \times 2 \rule[-4ex]{0pt}{8ex}$     &        $\begin{bmatrix}
 3 \times 3, &64 \\
  3 \times 3, &64 \\
   fc, & [2,64]
\end{bmatrix} \times 2 $       \\ \hline
conv3\_x                  &       $28 \times 28 \times 128 $            &      \rule{0pt}{4ex} $\begin{bmatrix}
 3 \times 3, &128 \\
  3 \times 3, &128 \\
  fc, & [4,128]
\end{bmatrix} \times 2 \rule[-4ex]{0pt}{8ex}$      &      \rule{0pt}{5ex} $\begin{bmatrix}
 3 \times 3, &128 \\
  3 \times 3, &128 \\
  fc, & [4,128]
\end{bmatrix} \times 2 \rule[-3ex]{0pt}{6ex}$        \\ \hline
conv4\_x                  &           $14\times 14 \times 256 $        &    \rule{0pt}{4ex} $\begin{bmatrix}
 3 \times 3, &256 \\
  3 \times 3, &256 \\
  fc, & [8,256]
\end{bmatrix} \times 2 \rule[-3ex]{0pt}{6ex}$       &         \rule{0pt}{4ex} $\begin{bmatrix}
 3 \times 3, &256 \\
  3 \times 3, &256 \\
  fc, & [8,256]
\end{bmatrix} \times 2 \rule[-3ex]{0pt}{6ex}$     \\ \hline
conv5\_x                  &         $7 \times 7 \times 512 $          &     \rule{0pt}{4ex} $\begin{bmatrix}
 3 \times 3, &512 \\
  3 \times 3, &512 \\
  fc, & [16,512]
\end{bmatrix} \times 2 \rule[-4ex]{0pt}{6ex}$      &        \rule{0pt}{5ex} $\begin{bmatrix}
 3 \times 3, &512 \\
  3 \times 3, &512  \\
  fc, & [16,512]
\end{bmatrix} \times 2 \rule[-3ex]{0pt}{6ex}$      \\ \hline
average pool              &          $1 \times 1 \times 64 $         &     $7 \times 7$  average pool    &         $7 \times 7$  average pool      \\ \hline
fully connected           &           $10$        &     $512 \times 10$ &          $512 \times 10$    \\ \hline
softmax                   &            $10$       &           &              
\end{tabular}
\label{tab18s}
}
\end{center}
\end{table*}

\clearpage

\begin{table*}[htbp]
\begin{center}
\caption{ResNet-50 and HQD ResNet-50 Structure}
\resizebox{0.8\textwidth}{!}{
\begin{tabular}{c|cc}
\hline
Output size                     & \multicolumn{1}{c|}{ResNet-50} &  HQD ResNet-50 \\ \hline
$112 \times 112$                & \multicolumn{2}{c}{conv, $7 \times 7$, 64, stride 2}                               \\ \hline
                                &\multicolumn{1}{c|}{ max pool, $3 \times 3$, stride 2 }&    HQD module,     $3 \times 3$                       \\ \cline{2-3} 
{$56 \times 56$}                &  \multicolumn{1}{c|}{\rule{0pt}{7ex}$\begin{bmatrix}
  \text{conv},& 1 \times 1 ,& 64\\
 \text{conv},& 3 \times 3 ,& 64 \\
 \text{conv},& 1 \times 1 ,& 256 
\end{bmatrix} \times 3$ }       &      $\begin{bmatrix}
  \text{conv},&1\times1,&64\\
 \text{conv},&3\times 3,&64 \\
 \text{conv},&1\times 1,&256 
\end{bmatrix} \times 3$    \\ \hline
$28 \times 28$                  &   \multicolumn{1}{c|}{\rule{0pt}{7ex} $\begin{bmatrix}
  \text{conv},&1\times1,&128\\
 \text{conv},&3\times 3,&128 \\
 \text{conv},&1\times 1,&512 
\end{bmatrix}\times4$}          &  $\begin{bmatrix}
  \text{conv},&1\times1,&128\\
 \text{conv},&3\times3,&128 \\
 \text{conv},&1\times1,&512  
\end{bmatrix} \times 4$             \\ \hline
$14 \times 14$                  & \multicolumn{1}{c|}{\rule{0pt}{7ex}  $\begin{bmatrix}
  \text{conv},& 1 \times 1 ,& 256\\
 \text{conv},& 3 \times 3 ,& 256 \\
 \text{conv},& 1 \times 1 ,& 1024 
\end{bmatrix} \times 6$  }      &  $\begin{bmatrix}
  \text{conv},&1\times1,&256\\
 \text{conv},&3\times3,&256 \\
 \text{conv},&1\times1,&1024  
\end{bmatrix} \times 6$             \\ \hline
$7 \times 7$                    & \multicolumn{1}{c|}{ \rule{0pt}{7ex}$\begin{bmatrix}
  \text{conv},& 1 \times 1 ,& 512\\
 \text{conv},& 3 \times 3 ,& 512 \\
 \text{conv},& 1 \times 1 ,& 2048 
\end{bmatrix} \times 3$     }   &  $\begin{bmatrix}
  \text{conv},&1\times1,&512\\
 \text{conv},&3\times3,&512 \\
 \text{conv},&1\times1,&2048
\end{bmatrix} \times 3$             \\ \hline
$1 \times 1$                    & \multicolumn{2}{c}{global average pool, 100-d $fc$, softmax}                     \\ \hline
\end{tabular}

}
\end{center}
\end{table*}

\begin{table*}[htbp]
\begin{center}
\caption{SE-ResNet-50 and HQD SE-ResNet-50 Structure}
\resizebox{0.8\textwidth}{!}{
\begin{tabular}{c|c|c}
\hline
Output size                     & SE-ResNet-50 & HQD SE-ResNet-50 \\ \hline
$112 \times 112$                & \multicolumn{2}{c}{conv, $7 \times 7$, 64, stride 2}                               \\ \hline
                                & \multicolumn{1}{c|}{max pooling, $3 \times 3$, stride 2}  & HQD module, $3 \times 3$                       \\ \cline{2-3} 
{$56 \times 56$}                & $\begin{bmatrix}
  \text{conv},&1\times1,&64\\
 \text{conv},&3\times 3,&64 \\
 \text{conv},&1\times 1,&256 \\
       fc,& [16,256] \rule[-1.2ex]{0pt}{3ex}& 
\end{bmatrix} \times 3$    & $\begin{bmatrix}
  \text{conv},&1\times1,&64\\
 \text{conv},&3\times 3,&64 \\
 \text{conv},&1\times 1,&256 \\
       fc,& [16,256] & 
\end{bmatrix} \times 3$      \\ \hline
$28 \times 28$                  & $\begin{bmatrix}
  \text{conv},&1\times1,&128\\
 \text{conv},&3\times3,&128 \\
 \text{conv},&1\times1,&512 \\
    fc,  & [32,512]\rule[-1.2ex]{0pt}{3ex} & 
\end{bmatrix} \times 4$             & $\begin{bmatrix}
  \text{conv},&1\times1,&128\\
 \text{conv},&3\times 3,&128 \\
 \text{conv},&1\times 1,&512 \\
    fc,   & [32,512] & 
\end{bmatrix} \times 4$        \\ \hline
$14 \times 14$                  & $\begin{bmatrix}
  \text{conv},&1\times1,&256\\
 \text{conv},&3\times3,&256 \\
 \text{conv},&1\times1,&1024 \\
       fc,& [64,1024] \rule[-1.2ex]{0pt}{3ex}& 
\end{bmatrix} \times 6$             & $\begin{bmatrix}
  \text{conv},&1\times1,&256\\
 \text{conv},&3\times 3,&256 \\
 \text{conv},&1\times 1,&1024 \\
     fc,  & [64,1024] & 
\end{bmatrix} \times 6$       \\ \hline
$7 \times 7$                    & $\begin{bmatrix}
  \text{conv},&1\times1,&512\\
 \text{conv},&3\times3,&512 \\
 \text{conv},&1\times1,&2048\\
     fc,  & [128,2048] \rule[-1.2ex]{0pt}{3ex} & 
\end{bmatrix} \times 3$             & $\begin{bmatrix}
  \text{conv},&1\times1,&512\\
 \text{conv},&3\times 3,&512 \\
 \text{conv},&1\times 1,&2048 \\
     fc,  & [128,2048] & 
\end{bmatrix} \times 3$      \\ \hline
$1 \times 1$                    & \multicolumn{2}{c}{global average pool, 100-d $fc$, softmax}                     \\ \hline
\end{tabular}

}
\end{center}
\end{table*}

\clearpage

\section{ResNeXt and HQD ResNeXt Structures}
\label{secxt}
\begin{table*}[htbp]
\begin{center}

\caption{ResNext-50 \cite{Xie2016AggregatedRT} and HQD ResNext-50 Structure}
\resizebox{1\textwidth}{!}{
\begin{tabular}{c|c|c}
\hline
Output size                     & ResNeXt-50 ($32 \times 4\text{d}$)& HQD ResNeXt-50 ($32 \times 4\text{d}$) \\ \hline
$112 \times 112$                & \multicolumn{2}{c}{conv, $7 \times 7$, 64, stride 2}                               \\ \hline
                                & \multicolumn{1}{c|}{max pooling, $3 \times 3$, stride 2}  & HQD module, $3 \times 3$                       \\ \cline{2-3} 
{$56 \times 56$}                & $\begin{bmatrix}
  \text{conv},&1\times1,&128& \\
 \text{conv},&3\times 3,&128 &, C =32 \\
 \text{conv},&1\times 1,&256&  \\
      
\end{bmatrix} \times 3$    & $\begin{bmatrix}
  \text{conv},&1\times1,&128 & \\
 \text{conv},&3\times 3,&128 &, C =32\\
 \text{conv},&1\times 1,&256 &   \rule[-1.2ex]{0pt}{3ex} \\

\end{bmatrix} \times 3$      \\ \hline
$28 \times 28$                  & $\begin{bmatrix}
  \text{conv},&1\times1,&256 &\\
 \text{conv},&3\times3,& 256 & , C =32\\
 \text{conv},&1\times1,&512  &
\end{bmatrix} \times 4$             & $\begin{bmatrix}
  \text{conv},&1\times1,&256 &\\
 \text{conv},&3\times 3,&256 & &, C =32\\
 \text{conv},&1\times 1,&512 & \rule[-1.2ex]{0pt}{3ex}
\end{bmatrix} \times 4$        \\ \hline
$14 \times 14$                  & $\begin{bmatrix}
  \text{conv},&1\times1,&512\\
 \text{conv},&3\times3,&512 & , C =32\\\
 \text{conv},&1\times1,&1024  \rule[-1.2ex]{0pt}{3ex}
\end{bmatrix} \times 6$             & $\begin{bmatrix}
  \text{conv},&1\times1,&512\\
 \text{conv},&3\times 3,&512 & , C =32\\\
 \text{conv},&1\times 1,&1024 \\
\end{bmatrix} \times 6$       \\ \hline
$7 \times 7$                    & $\begin{bmatrix}
  \text{conv},&1\times1,&1024\\
 \text{conv},&3\times3,& 1024 & , C =32\\\
 \text{conv},&1\times1,&2048 \rule[-1.2ex]{0pt}{3ex} \\
 
\end{bmatrix} \times 3$             & $\begin{bmatrix}
  \text{conv},&1\times1,&1024\\
 \text{conv},&3\times 3,&1024 & , C =32\\\
 \text{conv},&1\times 1,&2048 \\

\end{bmatrix} \times 3$      \\ \hline
$1 \times 1$                    & \multicolumn{2}{c}{global average pool, 100-d $fc$, softmax}                     \\ \hline
\end{tabular}
\label{tab50s}
}
\end{center}
\end{table*}

\clearpage

\section{Max Pooling and Quantum Downsampling Visualisation}
In this section, we present comparative feature maps for max pooling and quantum downsampling, offering an in-depth analysis across various datasets. The illustrations notably utilize \colorbox{red!90}{red circles} to highlight specific details that traditional max pooling overlooks, yet are retained through quantum downsampling.
\label{Ourmethod}

\begin{figure}[h]

\begin{center}
\centerline{\includegraphics[width=1\textwidth]{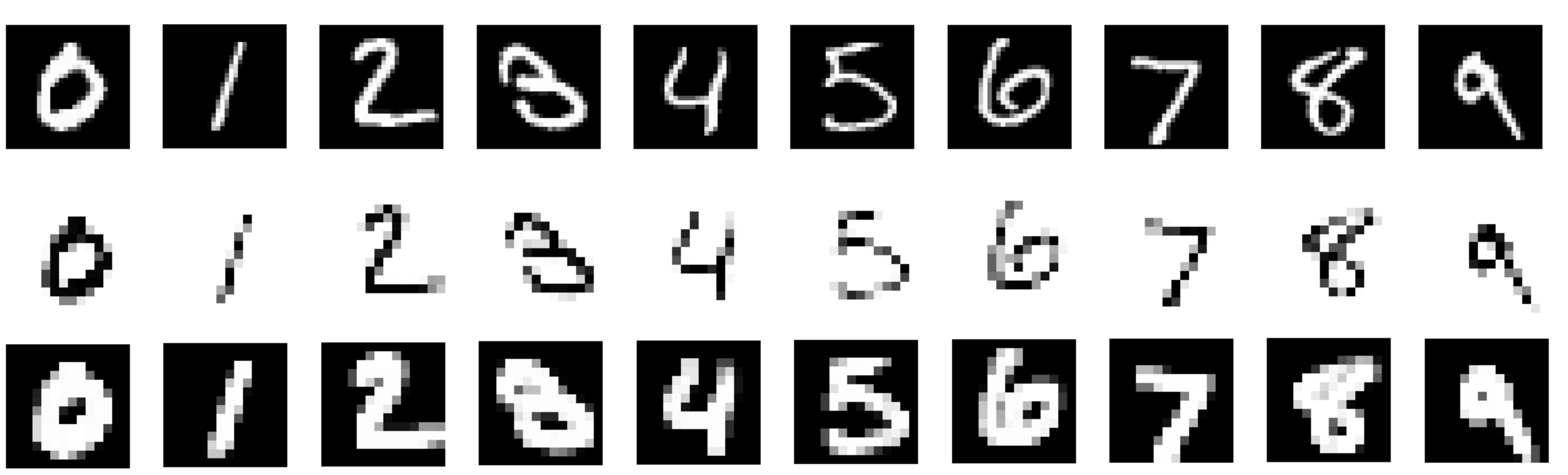}}
\caption{Visualization of images after different downsampling methods on \textbf{MNIST} from $0$ to $10$. Top: Original images. Middle: \textbf{Quantum Downsampling (Ours)}. Bottom: \textbf{Max pooling}. Kernel size is to be $3 \times 3$, stride is to be $2$, and padding is to be $1$.}

\vspace{-20 pt}
\end{center}
\end{figure}

\begin{figure*}[h]
\vskip 0.2in
\begin{center}
\centerline{\includegraphics[width=1\textwidth]{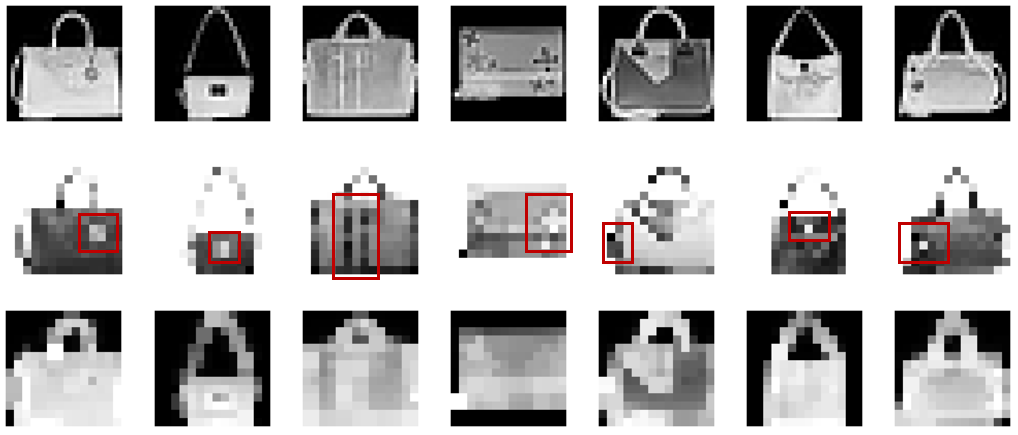}}
\caption{Visualization of images after different downsampling methods on \textbf{MNIST-Fasion}. Top: Original images. Middle: \textbf{Quantum Downsampling (Ours)}. Bottom: \textbf{Max pooling}. Both have kernel size to be $3 \times 3$, stride to $2$, and padding to $1$.}

\end{center}

\end{figure*}

\begin{figure*}[h]
\vskip 0.2in
\begin{center}
\centerline{\includegraphics[width=1\textwidth]{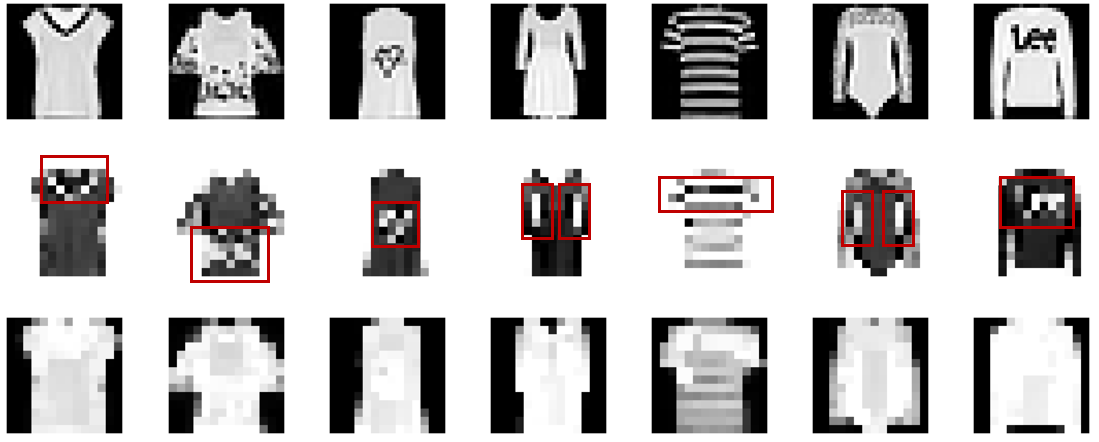}}
\caption{Visualization of images after different downsampling methods on \textbf{MNIST-Fasion}. Top: Original images. Middle: \textbf{Quantum Downsampling (Ours)}. Bottom: \textbf{Max pooling}. Both have kernel size to be $3 \times 3$, stride to $2$, and padding to $1$.}

\end{center}

\end{figure*}

\begin{figure*}[ht]

\begin{center}
\centerline{\includegraphics[width=1\textwidth]{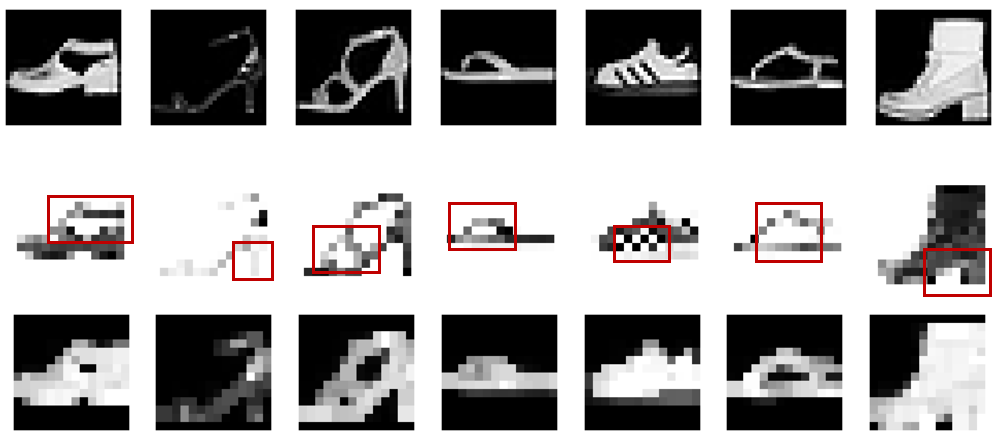}}
\caption{Visualization of images after different downsampling methods on \textbf{MNIST-Fasion}. Top: Original images. Middle: \textbf{Quantum Downsampling (Ours)}. Bottom: \textbf{Max pooling}. Kernel size is $3 \times 3$, stride is $2$, and padding is $1$.}

\end{center}
\end{figure*}

\begin{figure*}[h]
\vskip 0.2in
\begin{center}
\centerline{\includegraphics[width=1\textwidth]{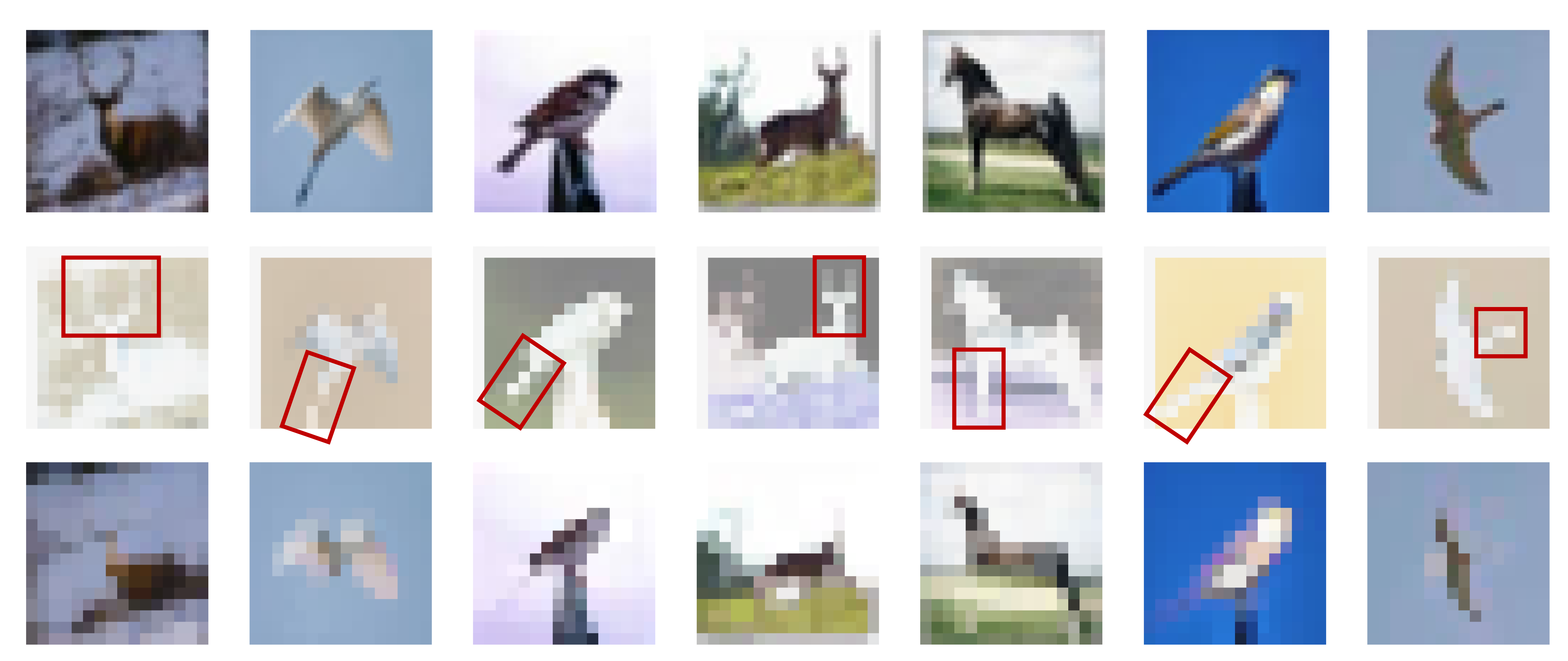}}
\caption{Visualization of images after different downsampling methods on \textbf{CIFAR-10}. Top: Original images. Middle: \textbf{Quantum Downsampling} (Ours). Bottom: \textbf{Max pooling}. Both have kernel size to be $3 \times 3$, stride to $2$, and padding to $1$.}

\end{center}

\end{figure*}

\begin{figure*}[h]
\vskip 0.2in
\begin{center}
\centerline{\includegraphics[width=1\textwidth]{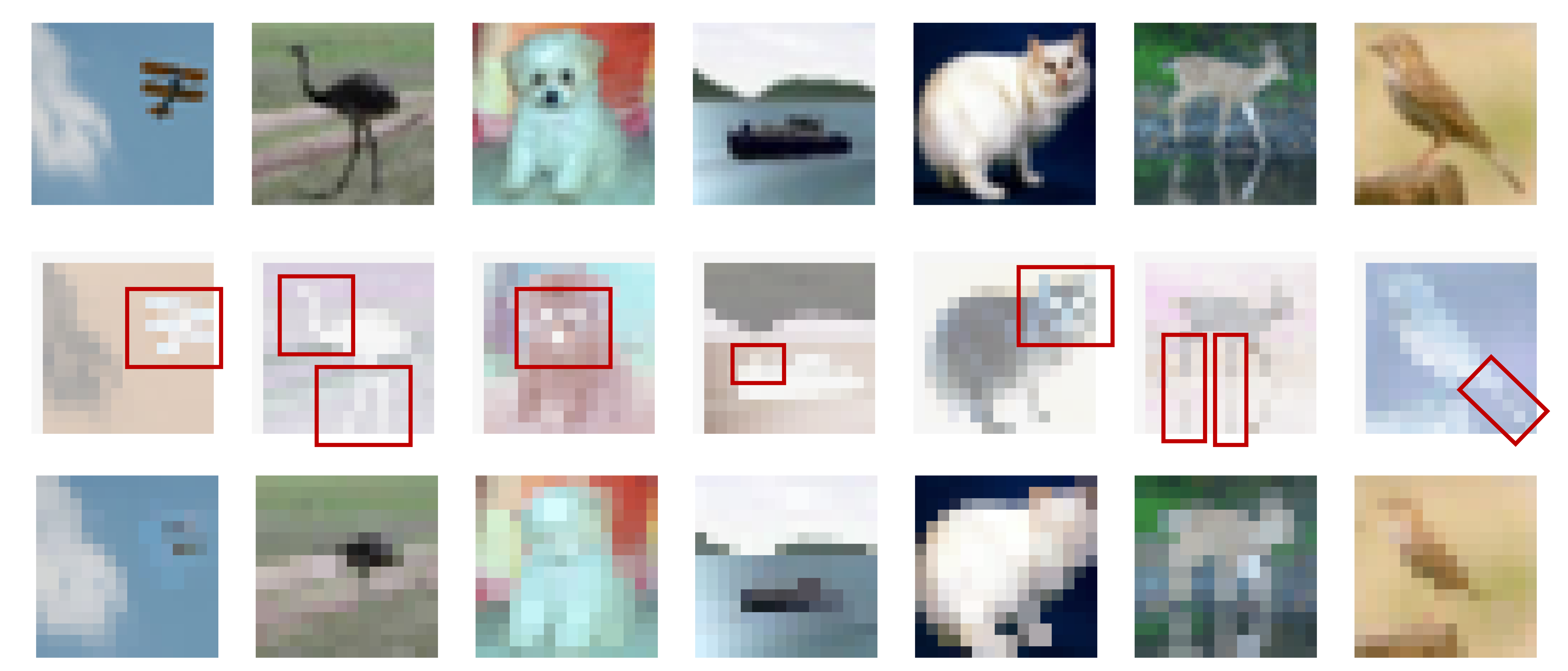}}
\caption{Visualization of images after different downsampling methods on \textbf{CIFAR-10}. Top: Original images. Middle: \textbf{Quantum Downsampling (Ours)}. Bottom: \textbf{Max pooling}. Both have kernel size to be $3 \times 3$, stride to $2$, and padding to $1$.}

\end{center}

\end{figure*}

\begin{figure*}[h]
\vskip 0.2in
\begin{center}
\centerline{\includegraphics[width=1\textwidth]{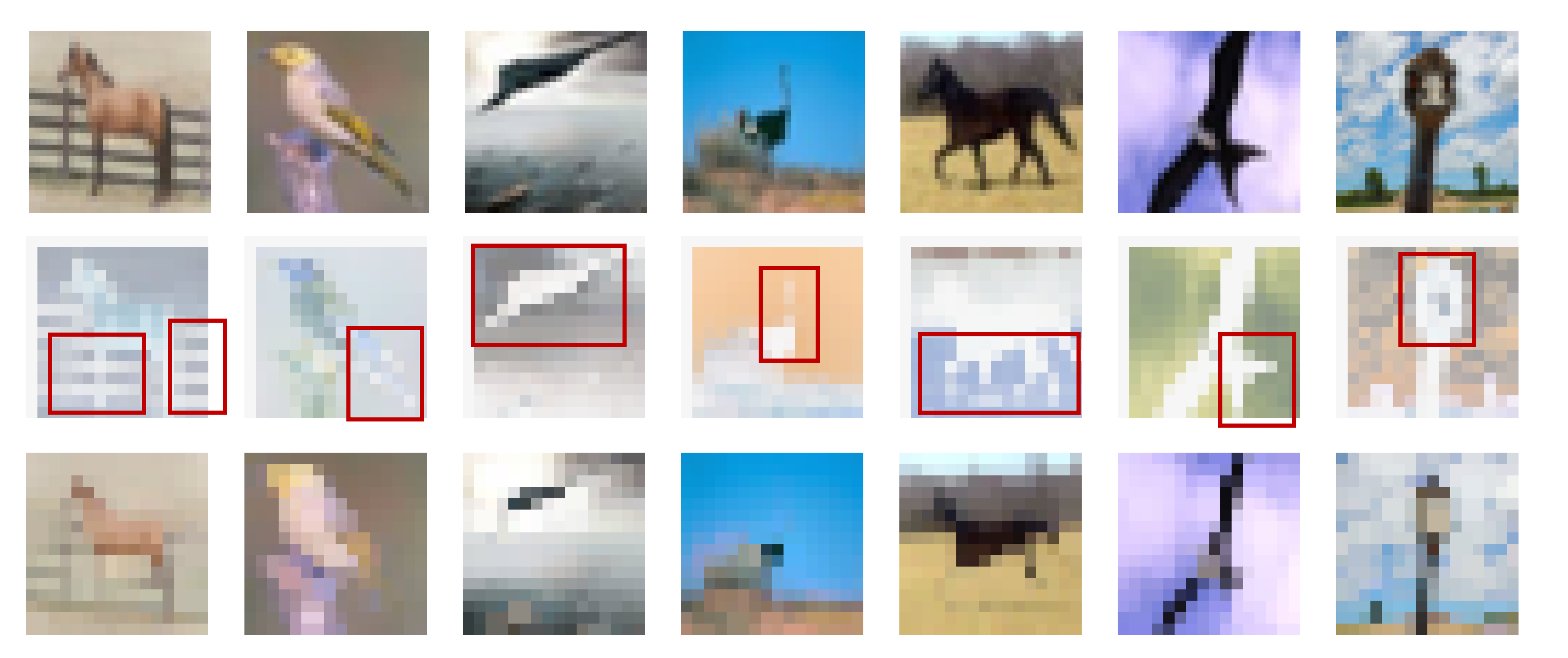}}
\caption{Visualization of images after different downsampling methods on \textbf{CIFAR-10}. Top: Original images. Middle: \textbf{Quantum Downsampling (Ours)}. Bottom: \textbf{Max pooling}. Both have kernel size to be $3 \times 3$, stride to $2$, and padding to $1$.}

\end{center}

\end{figure*}

\begin{figure*}[h]
\vskip 0.2in
\begin{center}
\centerline{\includegraphics[width=1\textwidth]{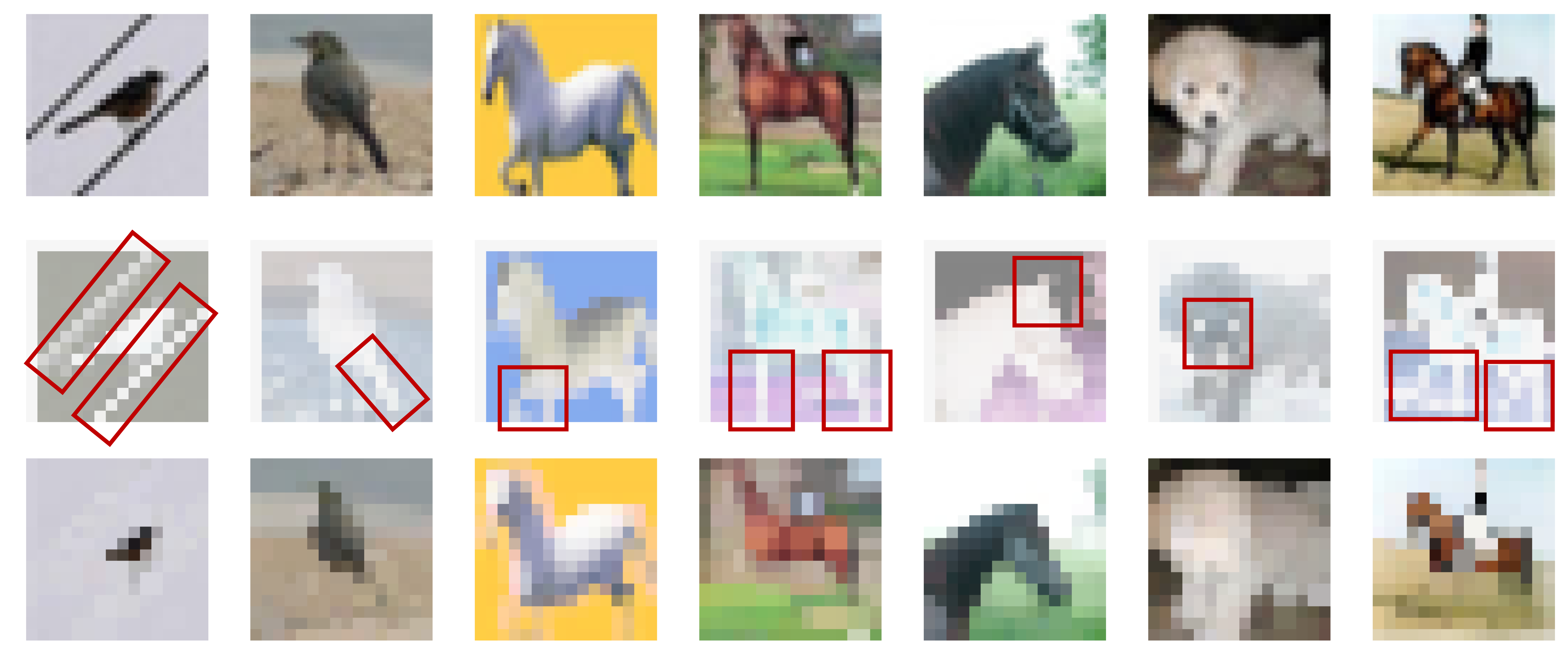}}
\caption{Visualization of images after different downsampling methods on \textbf{CIFAR-10}. Top: Original images. Middle: \textbf{Quantum Downsampling (Ours)}. Bottom: \textbf{Max pooling}. Both have kernel size to be $3 \times 3$, stride to $2$, and padding to $1$.}

\end{center}

\end{figure*}

\begin{figure*}[h]
\vskip 0.2in
\begin{center}
\centerline{\includegraphics[width=1\textwidth]{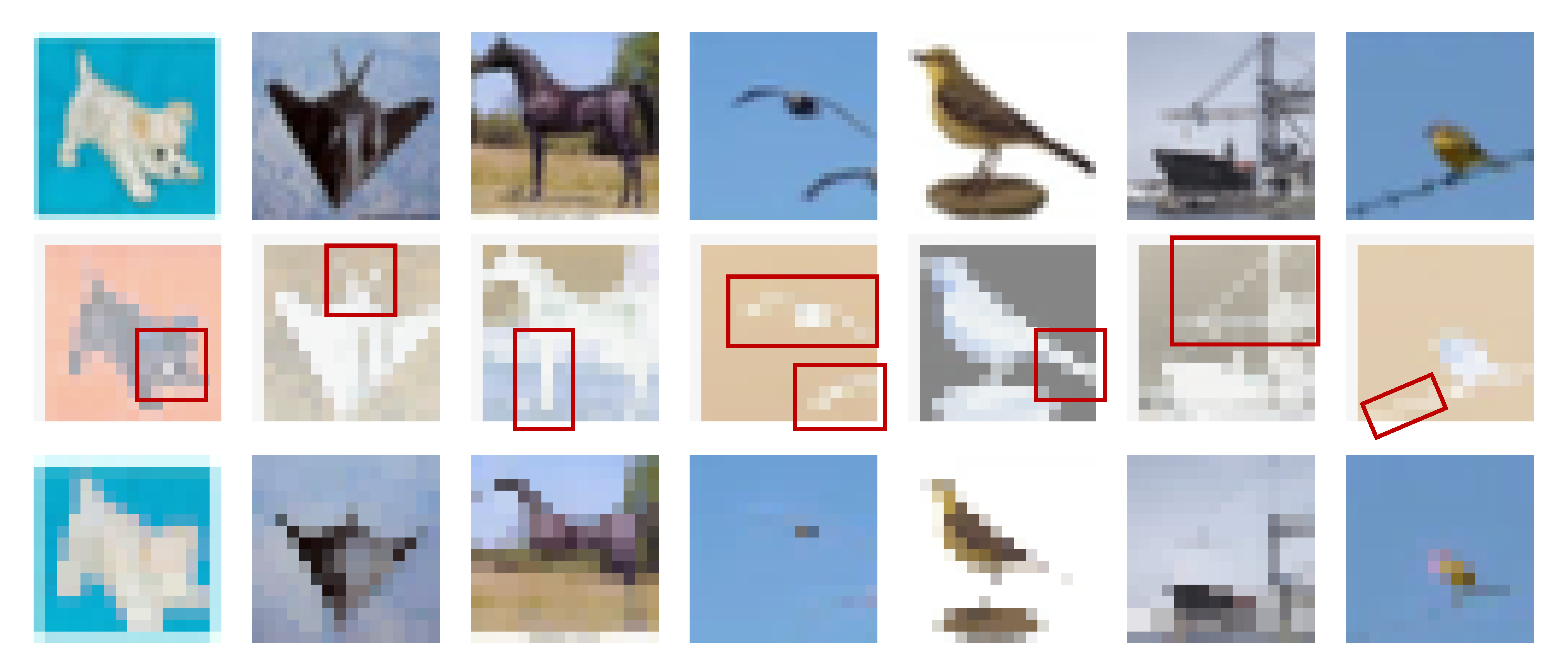}}
\caption{Visualization of images after different downsampling methods on \textbf{CIFAR-10}. Top: Original images. Middle: \textbf{Quantum Downsampling (Ours)}. Bottom: \textbf{Max pooling}. Both have kernel size to be $3 \times 3$, stride to $2$, and padding to $1$.}

\end{center}

\end{figure*}

\begin{figure*}[h]

\begin{center}
\centerline{\includegraphics[width=1\textwidth]{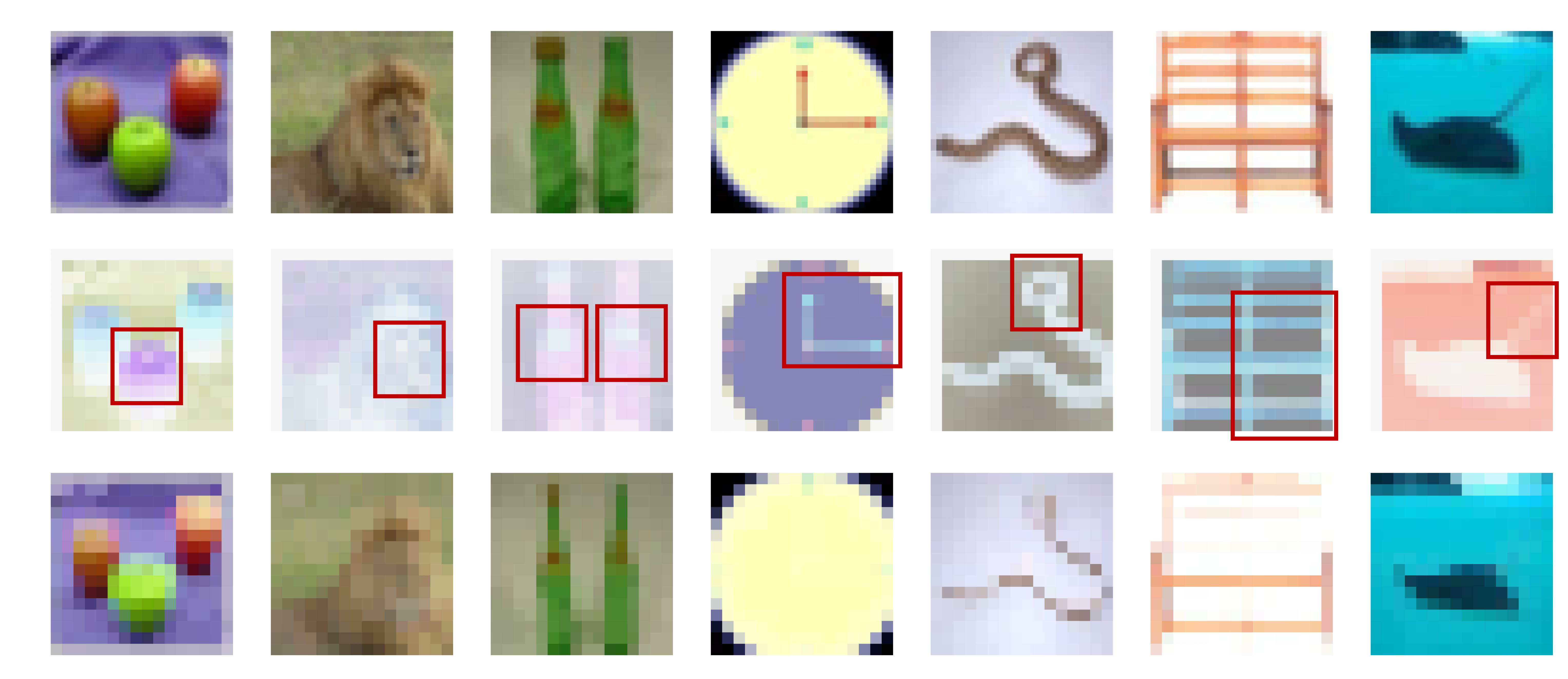}}
\caption{Visualization of images after different downsampling methods on \textbf{CIFAR-100}. Top: Original images. Middle: \textbf{Quantum Downsampling (Ours)}. Bottom: \textbf{Max pooling}. Both have kernel size to be $3 \times 3$, stride to $2$, and padding to $1$.}

\end{center}

\end{figure*}

\begin{figure*}[h]

\begin{center}
\centerline{\includegraphics[width=1\textwidth]{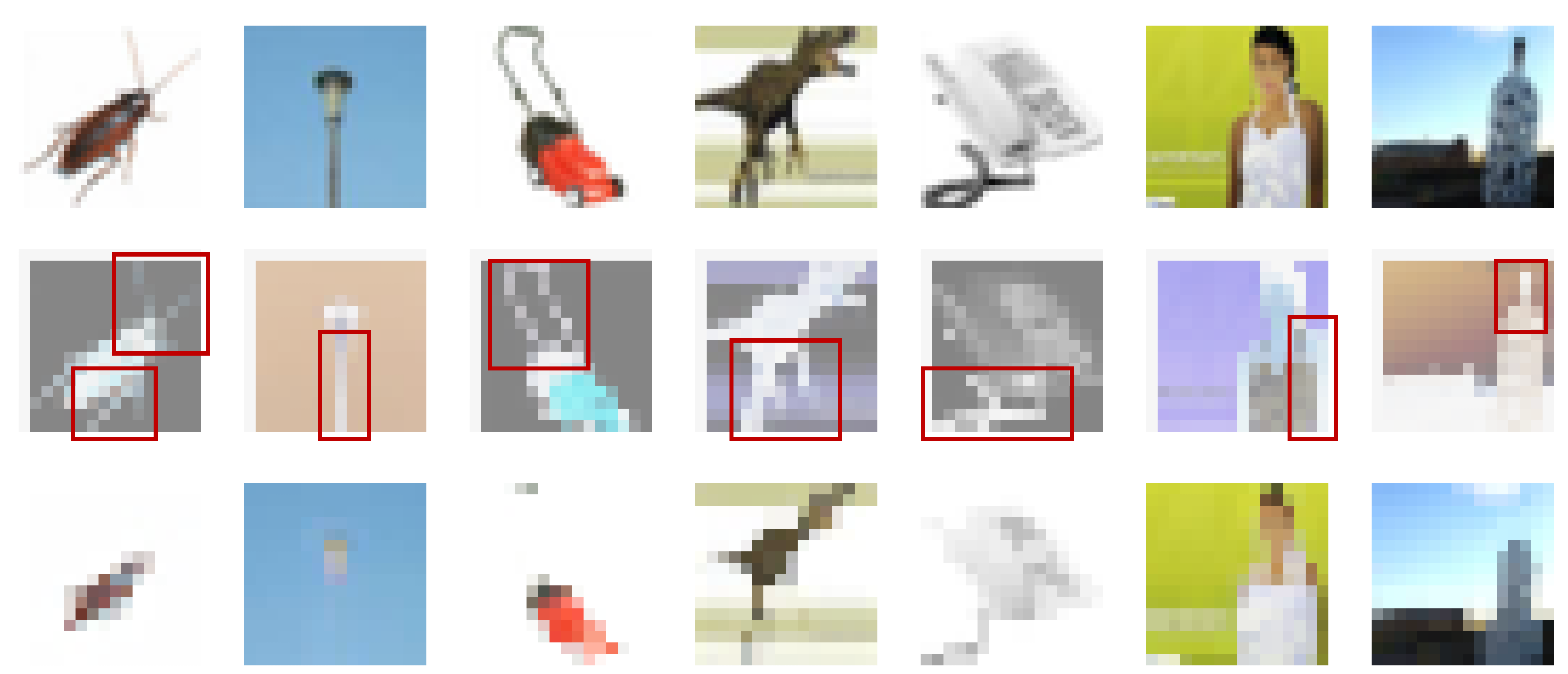}}
\caption{Visualization of images after different downsampling methods on \textbf{CIFAR-100}. Top: Original images. Middle: \textbf{Quantum Downsampling (Ours)}. Bottom: \textbf{Max pooling}. Both have kernel size to be $3 \times 3$, stride to $2$, and padding to $1$.}

\end{center}

\end{figure*}

\begin{figure*}[h]

\begin{center}
\centerline{\includegraphics[width=1\textwidth]{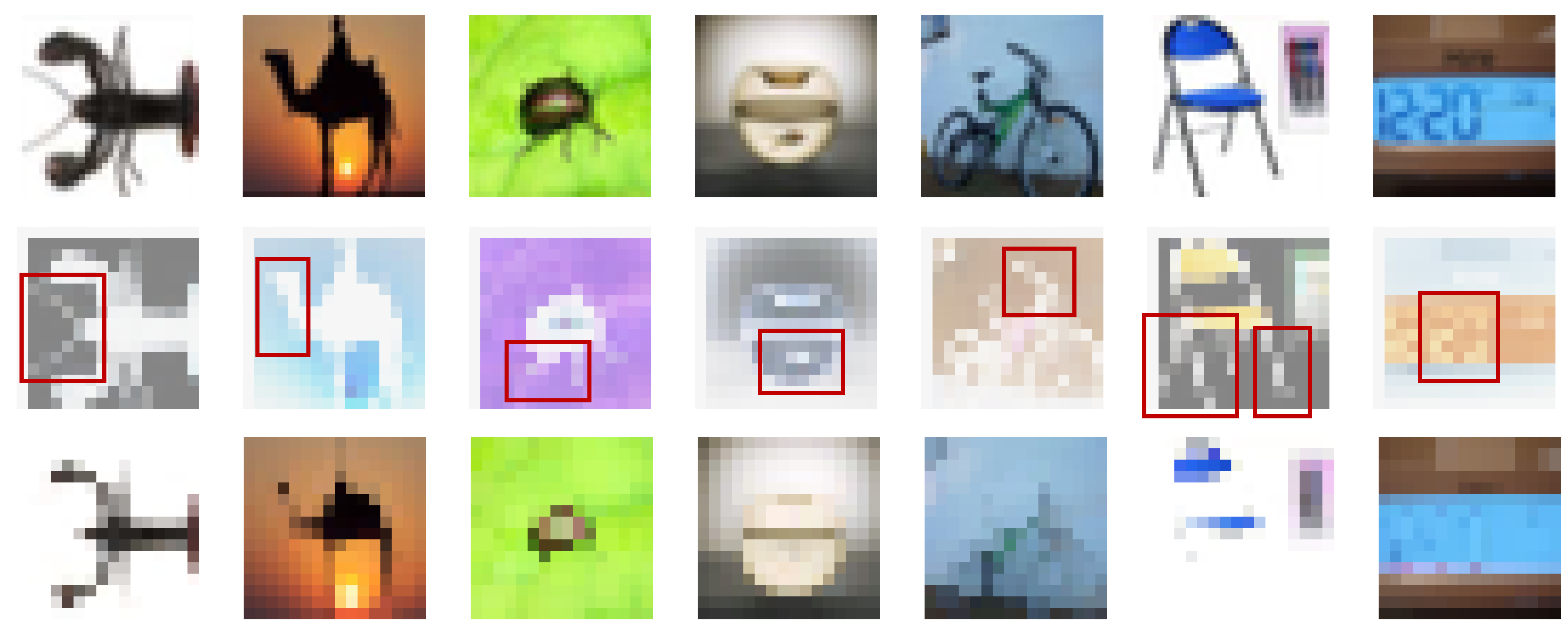}}
\caption{Visualization of images after different downsampling methods on \textbf{CIFAR-100}. Top: Original images. Middle: \textbf{Quantum Downsampling (Ours)}. Bottom: \textbf{Max pooling}. Both have kernel size to be $3 \times 3$, stride to $2$, and padding to $1$.}

\end{center}

\end{figure*}

\begin{figure*}[h]

\begin{center}
\centerline{\includegraphics[width=1\textwidth]{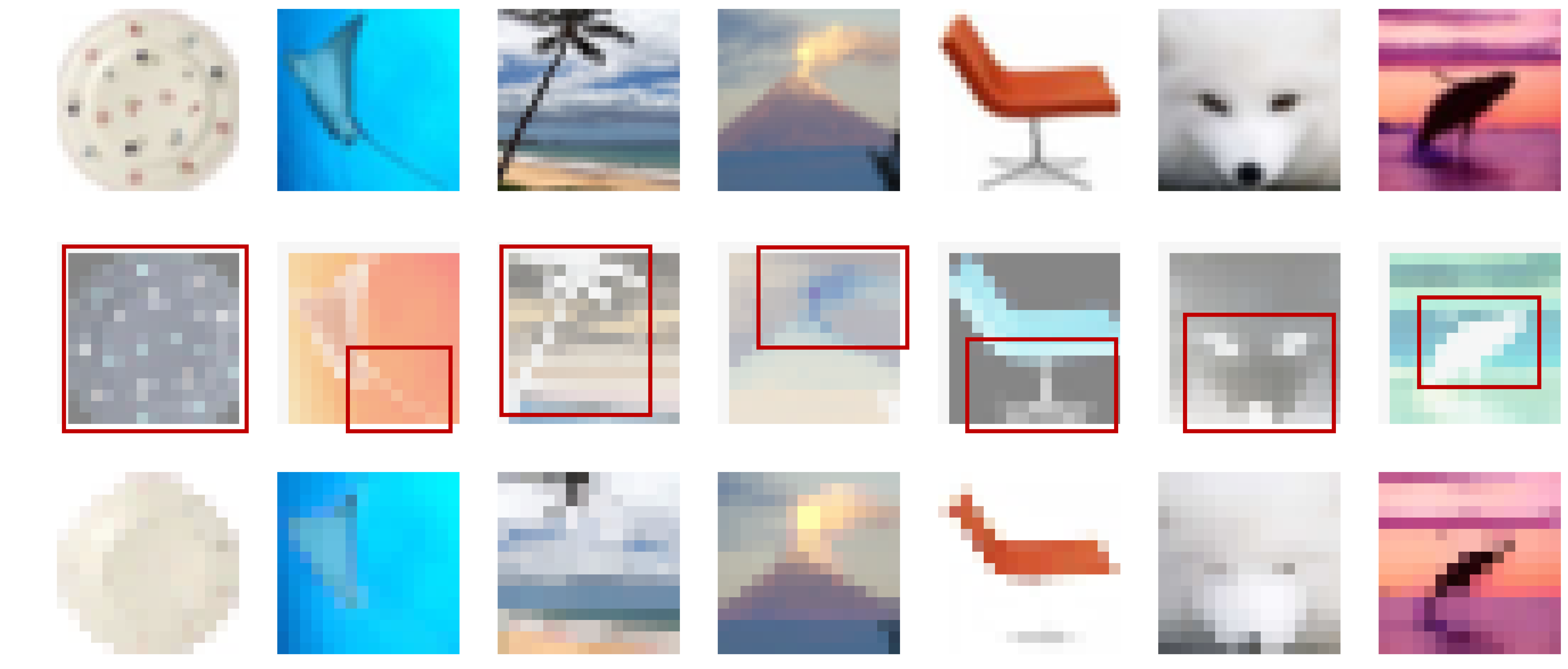}}
\caption{Visualization of images after different downsampling methods on \textbf{CIFAR-100}. Top: Original images. Middle: \textbf{Quantum Downsampling (Ours)}. Bottom: \textbf{Max pooling}. Both have kernel size to be $3 \times 3$, stride to $2$, and padding to $1$.}

\end{center}

\end{figure*}

\begin{figure*}[ht]

\begin{center}
    \centerline{\includegraphics[width=1\textwidth]{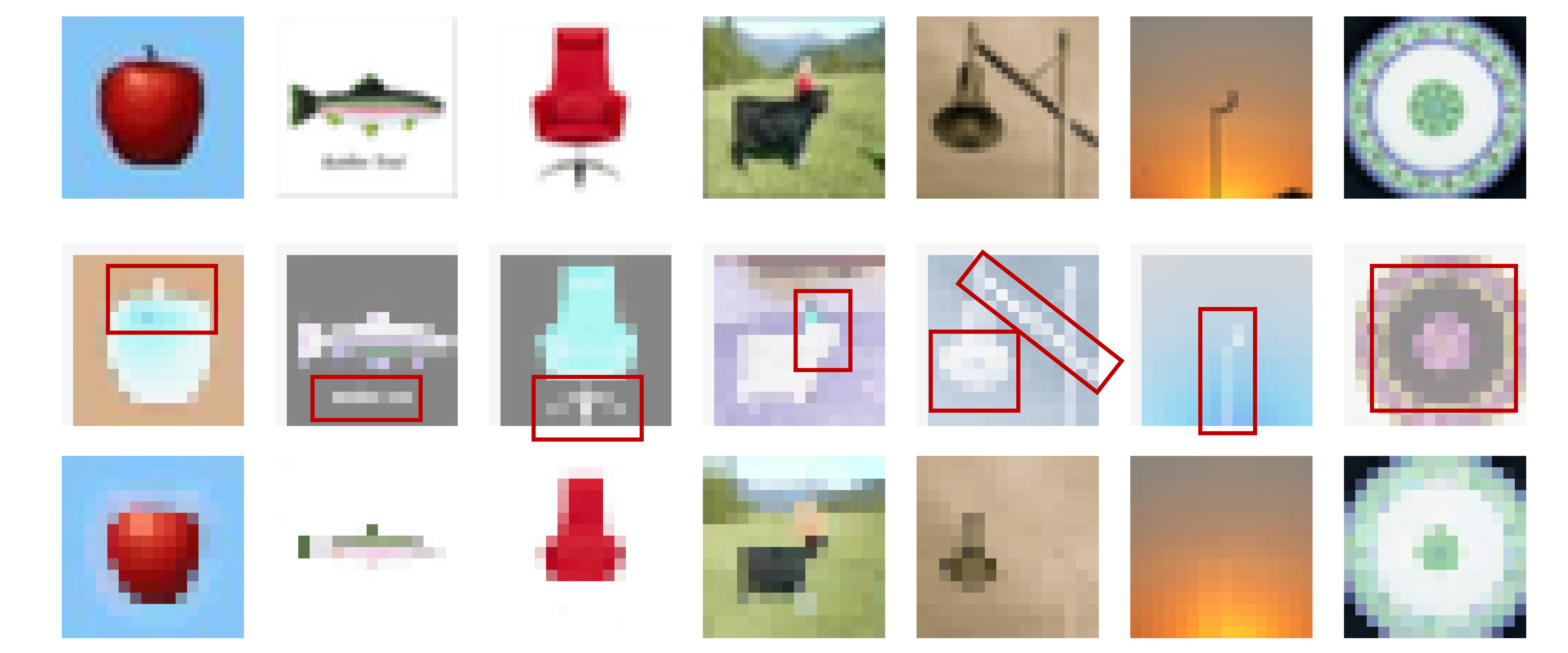}}
\caption{Visualization of images after different downsampling methods on \textbf{CIFAR-100}. Top: Original images. Middle: \textbf{Quantum Downsampling (Ours)}. Bottom: \textbf{Max pooling}. Kernel size is to be $3$, stride is to be $2$, and padding is to be $1$.}
\label{mnistviso}

\end{center}
\end{figure*}

\end{document}